\documentclass[a4paper,fleqn,usenatbib]{mnras}
\usepackage{newtxtext,newtxmath}
\usepackage[T1]{fontenc}
\usepackage{ae,aecompl}
\usepackage{natbib}
\usepackage{hyperref}
\usepackage{threeparttable}

\makeatletter
\def\@to{to}
\makeatother

\usepackage{graphicx}	
\usepackage{amsmath}	
\usepackage{amssymb}	

\newcommand{\CIVdblt}{\mathrm{C}\:\textsc{iv}~\lambda\lambda 1548, 1550}
\newcommand{\AlII}{\mathrm{Al}\:\textsc{ii}}
\newcommand{\CII}{\mathrm{C}\:\textsc{ii}}
\newcommand{\CIV}{\mathrm{C}\:\textsc{iv}}
\newcommand{\HI}{\mathrm{H}\:\textsc{i}}
\newcommand{\Lya}{\mathrm{Ly}\alpha}

\newcommand{\MgII}{\mathrm{Mg}\:\textsc{ii}}
\newcommand{\NV}{\mathrm{N}\:\textsc{v}}
\newcommand{\NI}{\mathrm{N}\:\textsc{i}}

\newcommand{\SiII}{\mathrm{Si}\:\textsc{ii}}
\newcommand{\SiIII}{\mathrm{Si}\:\textsc{iii}}
\newcommand{\SiIV}{\mathrm{Si}\:\textsc{iv}}
\newcommand{\kms}{\mathrm{km\,s}^{-1}}
\newcommand{\cmsq}{\mathrm{cm}^{-2}}
\newcommand{\cc}{\mathrm{cm}^{-3}}

\title[Cool gas in galaxy groups/clusters]{C IV absorbers tracing cool gas in dense galaxy group/cluster environments}

\author[Manuwal et al.]{Aditya Manuwal,$^{1}$\thanks{E-mail: aditya.manuwal@gmail.com}
Anand Narayanan,$^2$
Sowgat Muzahid,$^3$
Jane C. Charlton,$^4$
\newauthor
Vikram Khaire,$^5$
Hum Chand$^6$
\\
\\
$^{1,2}$Department of Earth and Space Sciences, Indian Institute of Space Science \& Technology, Thiruvananthapuram 695547, Kerala, INDIA\\
$^{3}$Leiden Observatory, Leiden University, PO Box 9513, 2300 RA, Leiden,  The Netherlands\\
$^{4}$The Pennsylvania State University, 413 Davey Lab, University Park, State College, PA 16802, USA\\
$^{5}$Department of Physics, University of California, Santa Barbara 93106, California, USA\\
$^{6}$Aryabhatta Research Institute of Observational Sciences (ARIES), Manora Peak, Nainital 263002, INDIA\\
}
\date{Accepted 2019 January 9. Received 2019 January 5; in original form 2018 June 7}

\pubyear{2019}

\begin{document}
\label{firstpage}
\pagerange{\pageref{firstpage}--\pageref{lastpage}}
\maketitle

\begin{abstract}
We present analysis on three intervening $\HI$-$\CIV$ absorption systems tracing gas within galaxy group/cluster environments, identified in the $HST$/COS far-UV spectra of the background quasars PG~$1148+549$ ($z_{abs}=0.00346$), SBS~$1122+594$ ($z_{abs}=0.00402$) and RXJ~$1230.8+0115$ ($z_{abs}=0.00574$). The ionization models are consistent with the origin of metal lines and $\HI$ from a cool and diffuse photoionized gas phase with $T \lesssim 4 \times 10^{4}$~K and $n_{\mathrm{H}} \lesssim 5 \times 10^{-4}~\cc$. The three absorbers have $89$, $51$ and $17$ galaxies detected within $1$~Mpc and $|\Delta v | < 600~\kms$. The RXJ~$1230.8+0115$ sightline traces the outskirt regions of the Virgo cluster where the absorber is found to have super-solar metallicity. The detection of metal lines along with $\HI$ has enabled us to confirm the presence of cool, diffuse gas possibly enriched by outflows and tidal interactions in environments with significant galaxy density.
\end{abstract}




\section{INTRODUCTION}

Progress in our understanding of the distribution and properties of baryons in the universe has required observations of diffuse gas outside of the luminous regions of galaxies. As simulations and observations have shown, the space between galaxies has remained the most dominant reservoir of baryons all through the history of the universe \citep[see reviews by][]{Rauch1998,Prochaska2009}. However, unlike at high redshifts ($z \gtrsim 3$) where a comprehensive understanding of these baryons is readily available through observations of the $\Lya$ forest \citep{Rauch1997,Weinberg1997}, the low redshift intergalactic baryons are a complex admixture of multiple density-temperature phases. These multiphase gas clouds, belonging to the circumgalactic (CGM) and the intergalactic medium (IGM), are a spinoff of the formation of structures in the universe such as galaxies, galaxy clusters and superclusters \citep{Persic1992,Cen1999,Cen2006,Dave2011,Valageas2002}. The CGM and IGM are further influenced by galactic scale processes such as mergers, gas accretion, and star formation driven outflows \citep{Heckman2001,Scannapieco2002,Strickland2004,Kobayashi2007,Rupke2011,Tripp2011,Muzahid2015extreme,Muratov2017,Wiseman2017}.

Much of the recent emphasis of UV absorption line studies has been in establishing the presence of shock-heated plasma of $T \sim 10^5-10^6$~K in the large scale environments surrounding galaxies \citep{Tripp2000,Narayanan2010,Narayanan2011,Danforth2011,Savage2011,Meiring2013,Savage2014,Pachat2016,Pachat2017}. The more tenuous baryons at $T \gtrsim 10^7$~K require emission and absorption measurements at X-ray wavelengths \citep{Buote2009,Fang2010,Williams2012,Ren2014}. These warm-hot gas phases are deemed important as they harbor as much as $50$\% of the cosmic baryon fraction, which is a factor of five more than the baryonic mass in galaxies \citep[e.g.,][]{Tripp2000,Dave2001}. 

Regions of galaxy overdensity such as groups and clusters also tend to possess substantial amounts of cool $T \sim 10^4 - 10^5$~K gas. Observations leading to an understanding of the properties of this cooler gas in cluster/group and associated large scale galaxy environments has been limited \citep{Rosenberg2003,Yoon2012,Burchett2016,Yoon2017,Muzahid2017,Burchett2018}. Besides being significant reservoirs of baryonic mass \citep{Gonzalez2007,Kravtsov2012,Emerick2015,Muzahid2017}, this gas phase could be a way to trace radiatively cooling flows in clusters, physical mechanisms like tidal interactions and gas stripping \citep[e.g.,][]{Jaffe2015}, as well as gas accretion through filaments of the cosmic web (e.g., \citet{Burns2010}).

In this paper, we present the detection and analysis of metal absorption lines associated with the Virgo cluster and two other clusters in its neighbouring environment. The absorption systems are detected in the archival $HST$/Cosmic Origins Spectrograph (COS) \citep{Green2012} spectra of three background quasars. In each case, there is detection of $\HI$ and $\CIV$ lines tracing $T \sim 10^4$~K gas in the respective galaxy overdensity regions. 

The $\HI$ associated with the Virgo cluster has been studied in great detail by \citet{Yoon2012}. Based on a sample of 25 $\Lya$ absorbers, the authors mapped the distribution and covering fraction of cooler ($T = 10^4 - 10^5$~K) gas within approximately one virial radius of the cluster. One of our sightlines (RXJ~$1230.8+0115$) overlaps with their sample. Whereas the Yoon et al. was exclusively about $\Lya$, the detection of $\CIV$ and other metal lines along with the $\HI$ has allowed us to estimate the density and gas temperature in the absorber. Additionally, the presence of metals has enabled us to establish the relative chemical abundances in the absorbing gas, which can be important for understanding the astrophysical origin of these absorbers. The two additional sightlines covered in this paper (PG~$1148+549$, SBS~$1122+594$) are within $15$~Mpc of M87, the giant elliptical galaxy that occupies the center of the Virgo cluster as known from diffuse X-ray emission studies \citep{Sarazin1986}. We explore the large-scale distribution of galaxies along both these sightlines at redshifts similar to Virgo where we find evidence for the presence of cool gas.  

Information on COS data is presented in Sec.~\ref{Sec2}. Description of the individual $\CIV$ absorbers and the line measurements are given in Sec.~\ref{Sec3}. Photoionization modelling of the absorbers and the physical properties derived from it are discussed in Sec.~\ref{Sec4}. In Sec.~\ref{Sec5}, the SDSS information on the large-scale distribution of galaxies proximate to each absorber is given, along with a discussion on its possible associations with intra-cluster gas as opposed to the CGM of nearby galaxies. Finally, we summarize the possible origins and the key modelling results for the three $\HI$ - $\CIV$ absorbers. Throughout, we use values of $H_0 = 69.6~\kms$ Mpc$^{-1}$, $\Omega_m = 0.286$ and $\Omega_\Lambda = 0.714$ given by \citet{Bennett2014}.

\section{DATA ANALYSIS} \label{Sec2}

This section describes the absorber and galaxy data used for this study. As part of a blind search to detect $\CIV$ absorbers in the low redshift universe, we identified four sightlines in the $HST$/COS Legacy Archive \footnote{https://archive.stsci.edu/hst/spectral\_legacy/} that probe the large scale environment around the Virgo cluster \citep[$z \sim 0.0036$][]{Ebeling1998}. Three sightlines (PG~$1148+549$, SBS~$1122+594$ and RXJ~$1230.8+0115$) were found to have detections of $\CIV$ at redshifts approximately coincident with the Virgo cluster, whereas PG~$1216+069$ had only a detection of $\HI$ with no associated metal lines in the COS spectrum at Virgo redshifts as seen in \citet{Yoon2012}. We therefore exclude PG~$1216+069$ from this study. However, \citet{Tripp2005} had detected some metal lines associated with this absorber using a higher resolution spectrum from Space Telescope Imaging Spectrograph (STIS) onboard $HST$. The archival COS spectra at medium resolution (FWHM $= 17 - 20~\kms$) were obtained with the G130M and G160M gratings as part of Prop IDs. $11741$ (PI. Todd Tripp), $11520$ (PI. James Green) and $11686$ (PI. Nahum Arav) respectively. The spectroscopic features of COS and its in-flight performance are explained in \citet{Green2012} and \citet{Osterman2011}. The coadded data for each sightline spans the wavelength interval $1150$~{\AA} to $1775$~{\AA}. The Nyquist sampled spectra have mean signal-to-noise ratios (per $17~\kms$ resolution element) of $17$, $12$, and $58$ for PG~$1148+549$, SBS~$1122+594$ and RXJ~$1230.8+0115$ respectively. 

Our search for $\CIV$ systems at $z > 0$ along these sightlines used the following criteria for establishing detections: (1) both $\lambda$1548 and $\lambda$1550 transitions of $\CIV$ should be covered by the COS spectra (2) $\CIV$~$1548$ should be detected at a significance of $\geq 3\sigma$, (3) for unsaturated lines, the equivalent width ratio for the doublet transitions should be approximately consistent with the expected value of $2:1$, and (4) the absorber has to be at $\Delta v > 5000~\kms$ from the emission redshift of the background QSO to exclude the absorbers potentially intrinsic to the quasar \citep[see e.g.,][]{Muzahid2013}. On the basis of these, we detected $\CIV$ absorbers at $z = 0.00346$, $z = 0.00402$, and $z = 0.00574$ towards PG~$1148+549$, SBS~$1122+594$, and RXJ~$1230.8+0115$ sightlines respectively, which are within $|\Delta v| \sim 1000~\kms$ of the Virgo cluster ($z_{cluster} = 0.0036$). The redshift of the absorbers were established based on wavelength of the pixel that showed peak optical depth in the $\CIV$~$1548$ line.

Low order polynomials were used to locally define the continuum after excluding obvious absorption features from the fitting region. Line measurements were carried out on the continuum normalized spectra through Voigt profile fitting and the apparent optical depth (AOD) method of \citet{Savage1991}. Profile fitting was done using the \textsc{VPFIT} routine \citep[version 10][]{Kim2007} by convolving the observed profile with the corresponding COS instrumental spread function from \citet{Kriss2011}. 

Information on galaxies was obtained from the Data Release 14 of the Sloan Digitial Sky Survey (SDSS) archive \citep{Abolfathi2017}. At $z = 0.004$, the SDSS galaxy spectroscopic data is $90$\% complete down to $r < 17.8$ \citep{Strauss2002}, corresponding to $L \sim 0.001L^*$ at $z = 0.04$ \citep{Blanton2003galaxy}, which is adequate for gathering a full understanding of the galaxy distribution near the absorbers.

\section{DESCRIPTION OF ABSORBERS} \label{Sec3}

\subsection{The $\mathbf{z_{abs}=0.00346}$ absorber towards PG~$\mathbf{1148+549}$} \label{Danforth}

\begin{figure*}
    \includegraphics[height=630pt,trim=1cm 1.3cm 3cm 5.5cm,clip=true]{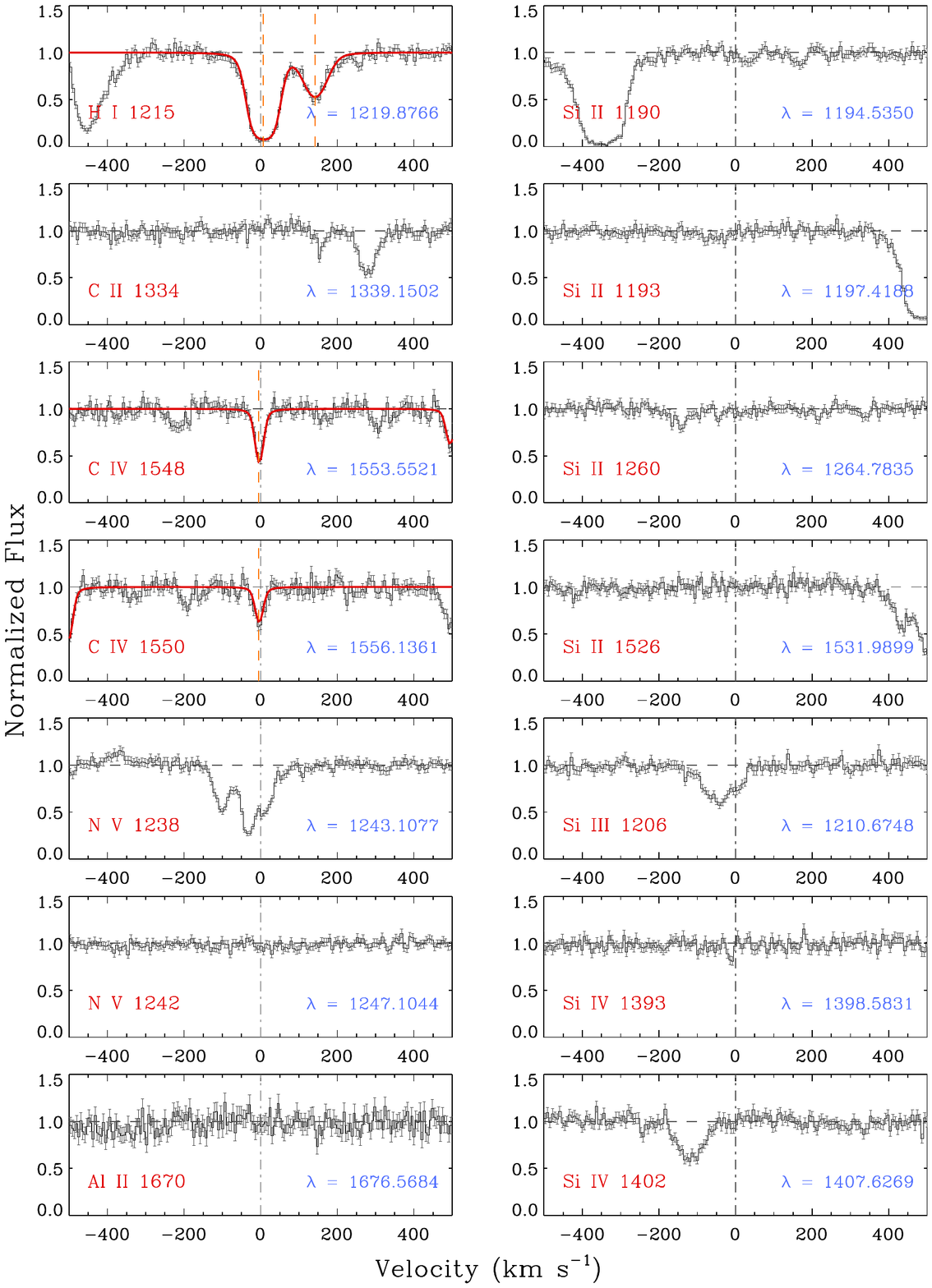}
    \caption{System plot of the $z_{abs} = 0.00346$ $\CIV$ absorber towards PG~$1148+549$ with free-fit of the saturated $\HI$. The zero velocity corresponds to the redshift of the absorber derived from wavelength of the pixel that shows maximum optical depth in the $\CIV$~$1548$ line indicated by the \textit{dashed-dot} vertical line in each panel. The Y-axis is continuum normalized flux. The error bars represent $1\sigma$ uncertainty in flux values. The \textit{red} curves overplotted on the spectra represent the best-fit Voigt profiles. The output parameters of profile fitting are listed in Table~\ref{tab1}. The \textit{dashed} vertical lines mark the line centroid given by the fitting routine.}
    \label{1}
\end{figure*}

\begin{figure*}
	    \includegraphics[width=245pt,height=151pt,trim=0.42cm 3.25cm 0.4cm 1.85cm,clip=true]{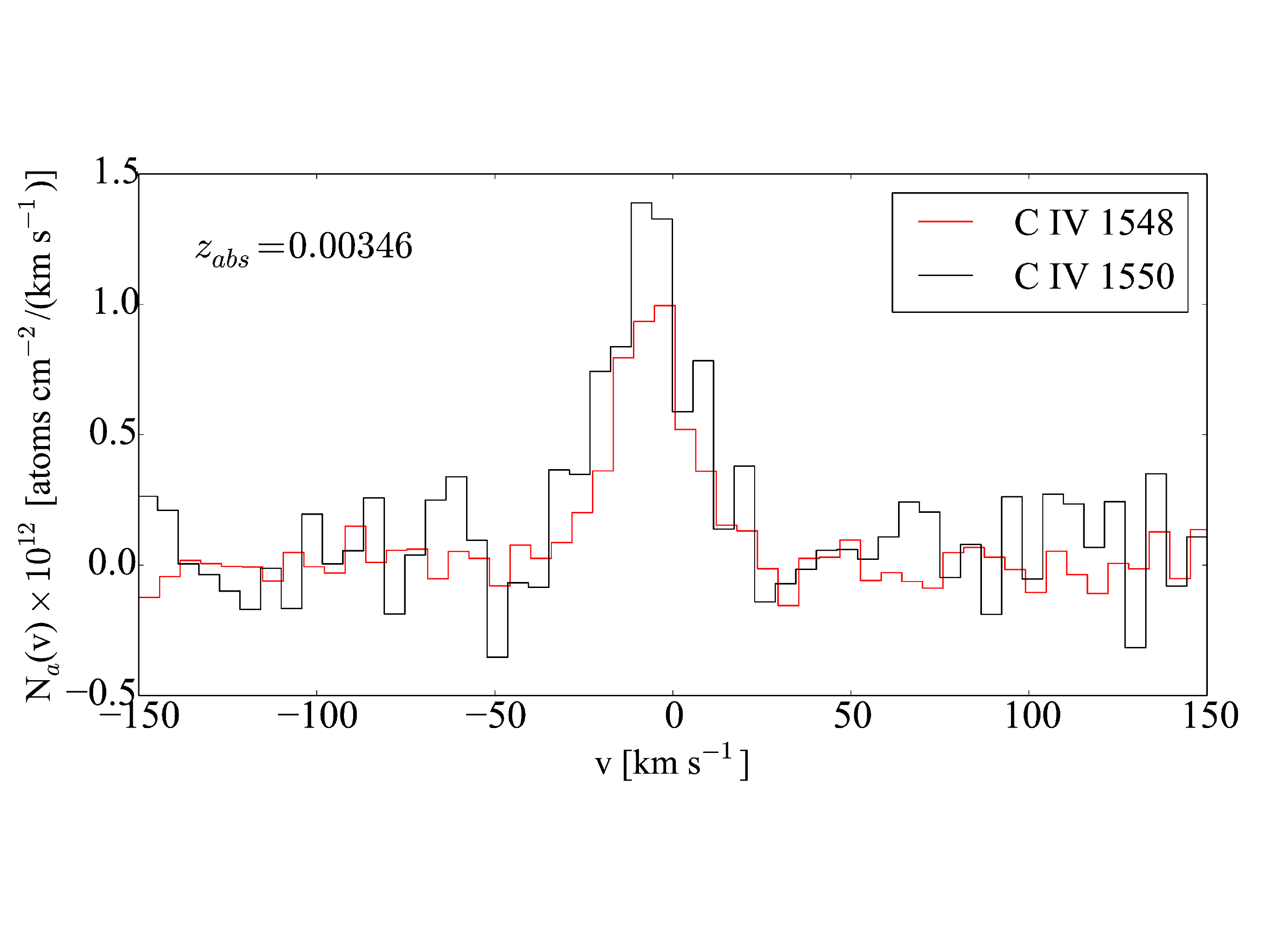} \quad 
        \includegraphics[width=245pt,height=135pt,trim=0.26cm 3cm 0cm 3.42cm,clip=true]{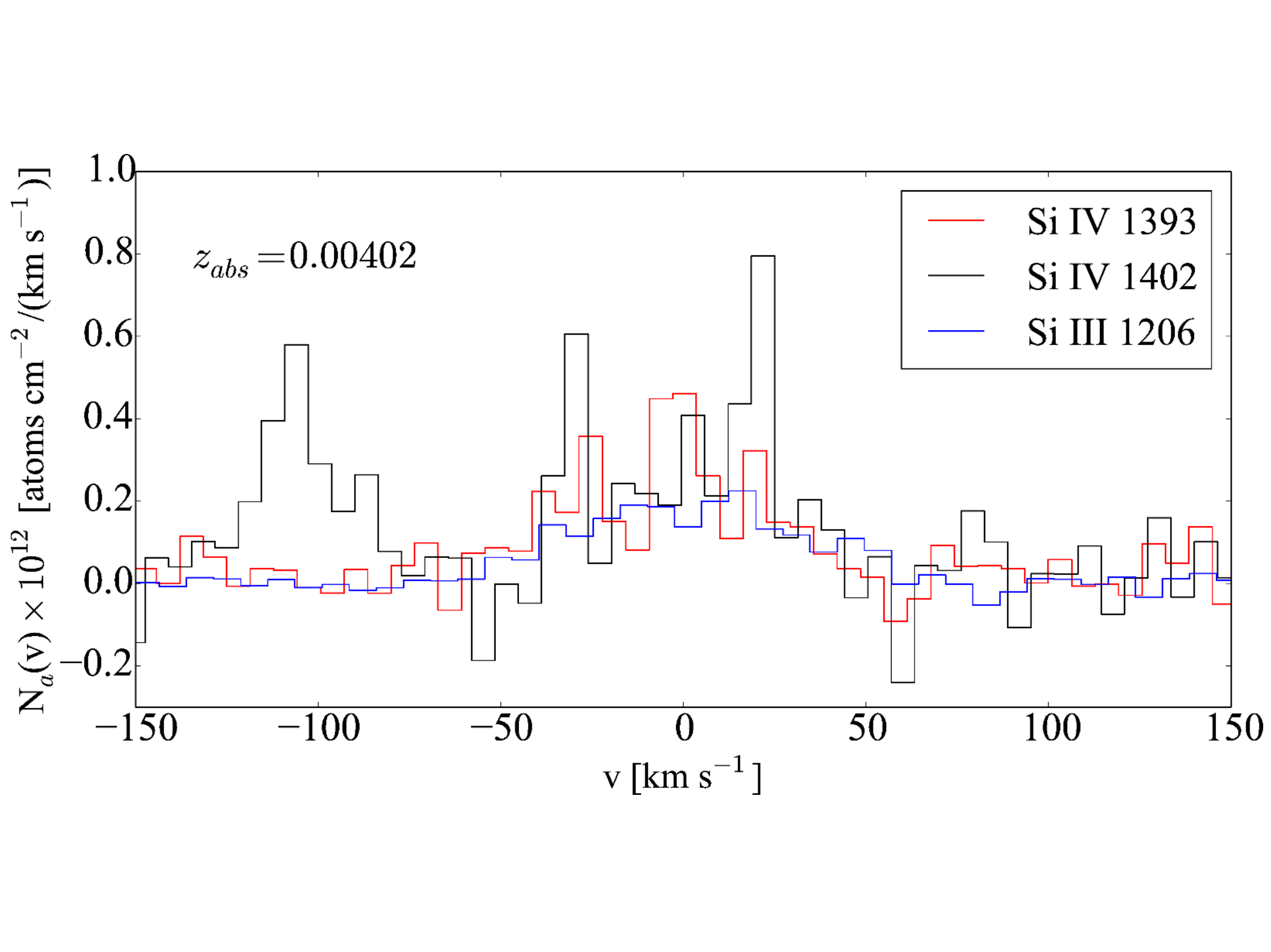} \quad 
        \includegraphics[width=240pt,height=133pt,trim=-0.5cm 3.3cm 0.4cm 3.55cm,clip=true]{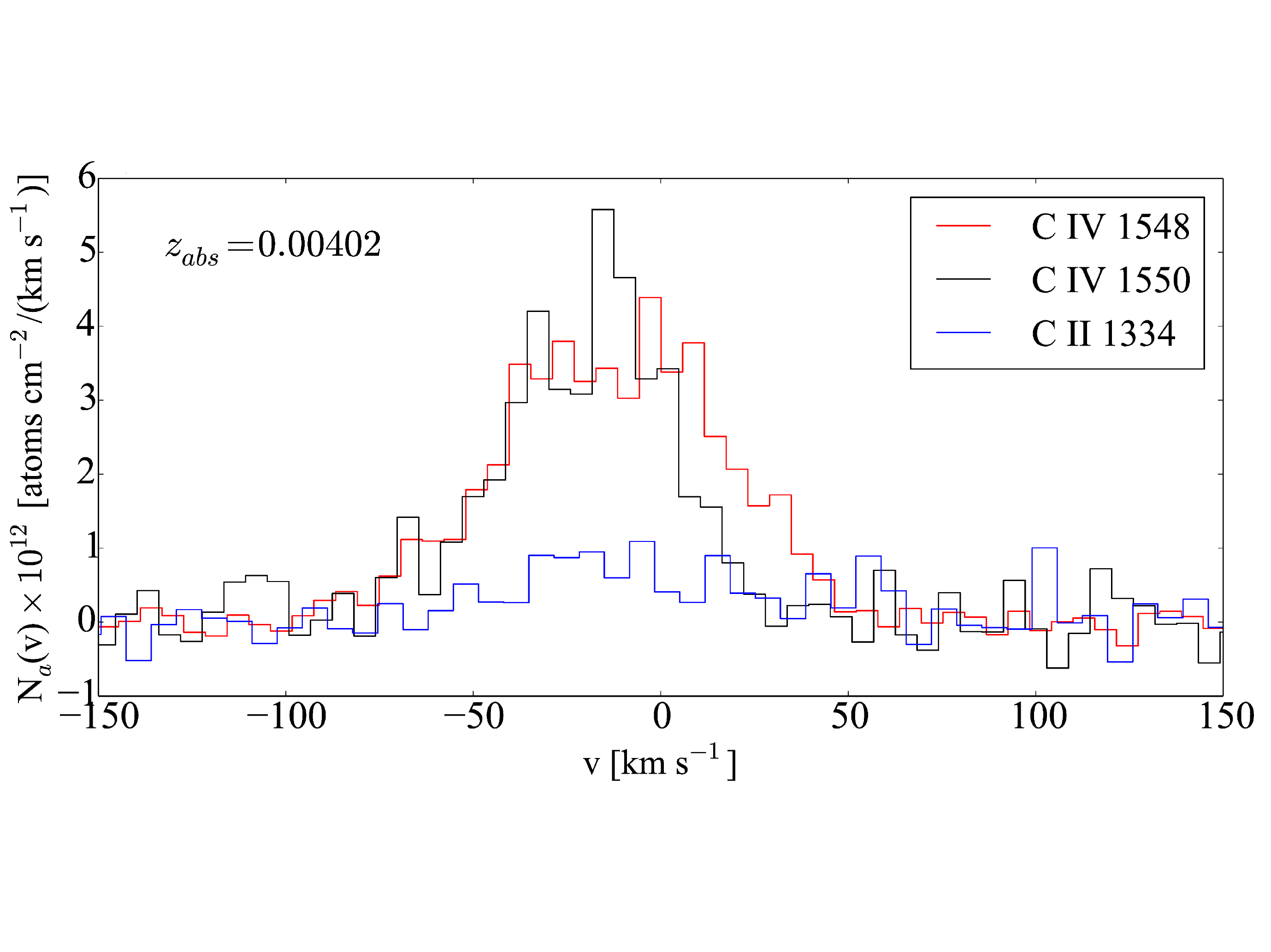} \quad 
        \includegraphics[width=249pt,height=133pt,trim=-0.5cm 3cm 0.19cm 3.38cm,clip=true]{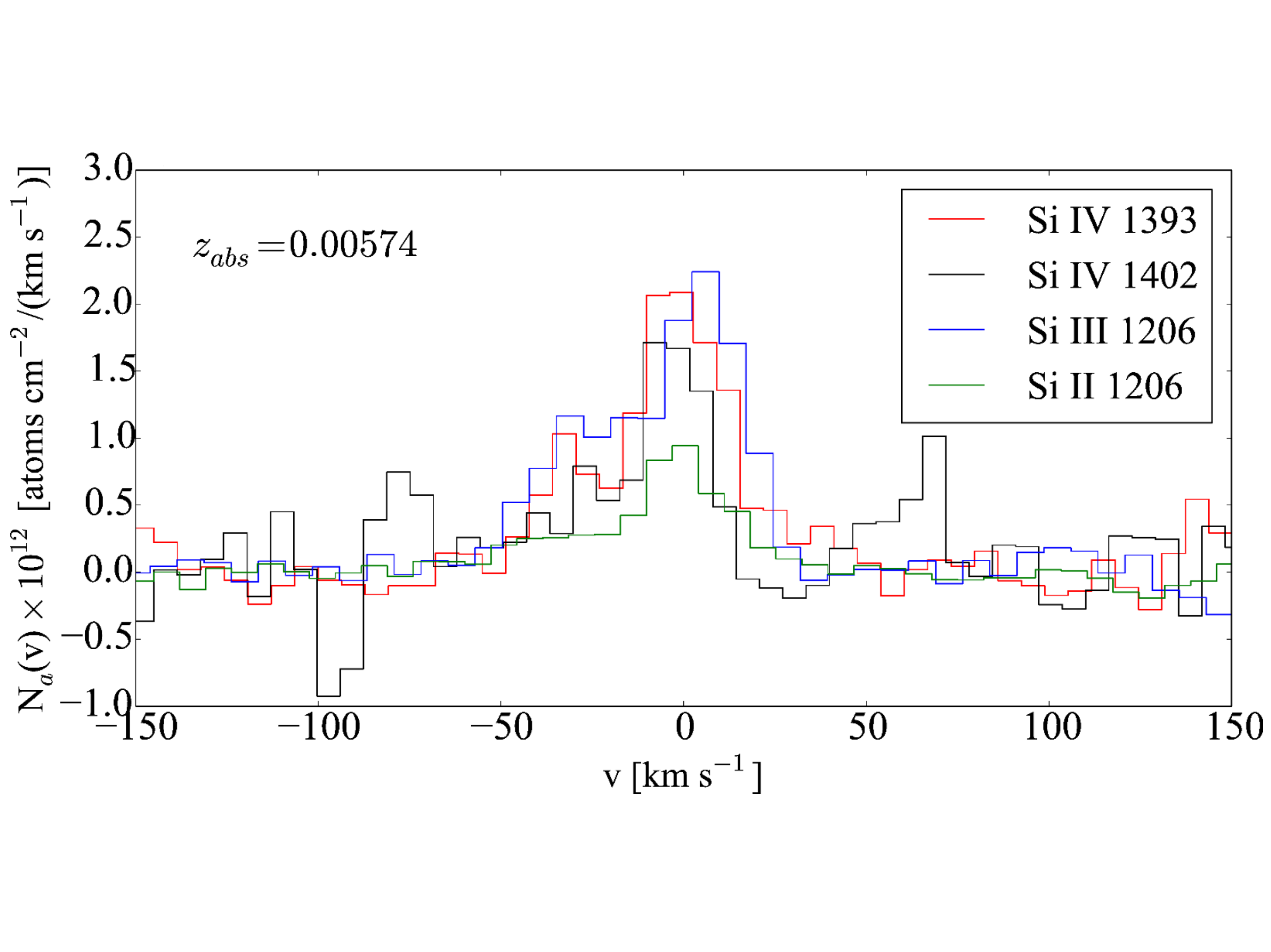} \quad 
        \includegraphics[width=245pt,height=131pt,trim=0.2cm 3.6cm 0.4cm 3.3cm,clip=true]{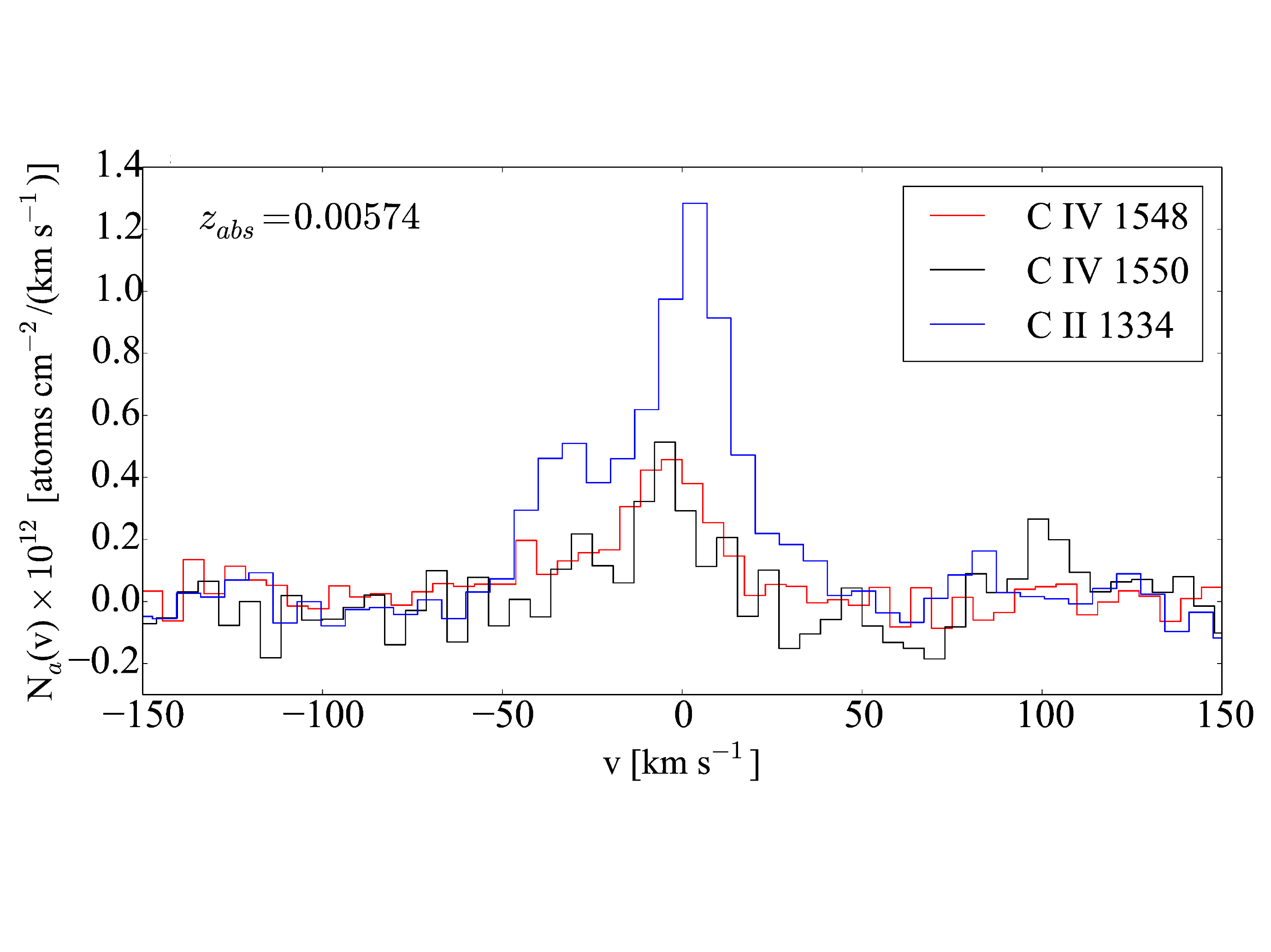}
    \caption{The figures show the apparent column density ($N_a(v)$) profiles for the C and Si transitions in the three absorbers. The absorber redshifts are indicated in the respective panels. The $N_a(v)$ comparison for the $z=0.00346$ system shows mild levels of unresolved saturation in the $\CIV$~$1548$ line core, whereas for the $z = 0.00402$ and $z = 0.00574$ systems, there is little evidence for line saturation. The excess $N_a(v)$ seen in the $z = 0.00402$ $\CIV$~$1548$ line between $+5 \leq \Delta v \leq +45~\kms$ indicates unidentified contamination. The contaminating pixels are excluded during simultaneous Voigt profile fitting of the $\CIVdblt$ lines. In the $z = 0.00574$ system, the kinematic resemblance of $\CII$ with the $\CIV$ lines is suggestive of the the two ions tracing a single gas phase.}
    \label{2}
\end{figure*}

\begin{table}
\caption{Line measurements for the $z = 0.00346$ absorber towards PG~$1148+549$.} 
\begin{center}
\begin{threeparttable}
\begin{tabular}{lccc}
\hline
Line     &	$W_r$   	   &       log~$[N_a~(\cmsq)]$	&	$[-v, +v]$     \\ 
         &      (m{\AA})           &                            &          ($\kms$)      \\  
\hline
$\HI$~$1215$		&   $378~{\pm}~27$      &	$>14.18$			& [-125, 66] \\
                        &   $170~{\pm}~26$      &	$13.58~{\pm}~0.02$	        & [66, 221] \\
$\CII$~$1334$		&   $< 18$	        &	$< 12.9$	        & [-45, 20] \\
$\AlII$~$1670$		&   $< 33$	        &	$< 12.0$ 	        & [-45, 20] \\
$\SiII$~$1190$          &   $< 15$		&	$< 12.6$	        & [-45, 20] \\
$\SiII$~$1193$          &   $< 15$		&	$< 12.3$	        & [-45, 20] \\
$\SiII$~$1260$          &   $< 15$		&	$< 11.9$	        & [-45, 20] \\
$\SiII$~$1304$          &   $< 21$		&	$< 13.2$	        & [-45, 20] \\
$\SiII$~$1526$          &   $< 21$		&	$< 12.9$	        & [-45, 20] \\
$\CIV$~$1548$           &   $81~{\pm}~6$	&	$13.41~{\pm}~0.07$	& [-45, 20] \\
$\CIV$~$1550$ 	        &   $70~{\pm}~7$       &	$13.62~{\pm}~0.09$      & [-45, 20] \\
$\NV$~$1242$            &   $< 15$              &	$< 13.1$	        & [-45, 20] \\
$\SiIII$~$1206$		&   $< 77$	        & 	$< 12.7$	        & [-45, 20] \\	
$\SiIV$~$1393$		&   $< 18$	        &	$< 12.3$	        & [-45, 20] \\
$\SiIV$~$1402$		&   $< 15$	        &	$< 12.3$	        & [-8, 20] \\
Line     &	$v$      &       log~$[N~(\cmsq)]$	&    $b$     \\
         &     ($\kms$)    &                              &  ($\kms$)    \\
$\HI$~$1215$	        &   $6~{\pm}~2$		&	$(14.60 - 17.61)$	 &	$(10 - 35)$	\\
                        &   $142~{\pm}~3$	&	$13.50~{\pm}~0.08$	 &	$29~{\pm}~5$	\\
$\CIV$~$1548-1550$	&   $-5~{\pm}~1$	&	$13.59~{\pm}~0.03$	 &	$10~{\pm}~2$	\\
\hline
\end{tabular}
\label{tab1}
\begin{tablenotes}
\item[] The \textit{top} portion of the table lists the rest-frame equivalent widths and integrated apparent column densities for the various species. The \textit{lower} portion lists the line parameters obtained from Voigt profile fits. Except for $\CIV$ and $\Lya$ all other lines are non-detections at the $\geq 3\sigma$ significance level. The $\SiIII$~$1206$ is contaminated by absorption unrelated to the system, yielding an uncertain upper limit on the equivalent width and column density.
\end{tablenotes}
\end{threeparttable}
\end{center}
\end{table}

The system plot for the absorber is shown in Figure~\ref{1}, and the AOD and profile fit measurements are listed in Table~\ref{tab1}. The absorber is detected only in $\CIV$ and $\HI$. The low ($\CII$, $\AlII$, $\SiII$) and intermediate ionization lines ($\SiIII$, $\SiIV$, and $\NV$) are non-detections. The comparison between the apparent column density profiles of the $\CIV$ doublets suggest only small amounts of unresolved saturation at the $\CIV$~$1548$ line core (see Figure~\ref{2}). The $\CIVdblt$ lines have identical kinematic profiles well represented by a single component fit. The integrated apparent column density for either line is also consistent with the values obtained from Voigt profile fitting. The $b(\CIV)$ gives an upper limit on the temperature of the gas as $T < 7 \times 10^{4}$~K. The $\Lya$ absorption is fitted with two components with one of them being coincident in velocity with the $\CIV$ to within one resolution element. The $\Lya$ line coincident with $\CIV$ is saturated. Voigt profile modelling therefore does not offer a unique solution to this component. The range of values for $N$ and $b$ that can yield satisfactory fits to the saturated $\Lya$ component can be estimated by varying the $b$-value of $\HI$ within the plausible range allowed by the narrow $\CIV$ line width. Assuming a pure thermal broadening scenario ($b(\HI) = 3.452 \times b(\CIV)$) yields an upper limit on the $b$-value, and pure non-thermal broadening ($b(\HI) = b(\CIV)$) gives the lower limit. The profile models from these two limiting $b$-values of $10 - 35~\kms$ yield good fits to the saturated $\HI$ component, with a corresponding wide column density range of $14.60 \leq \log~[N(\HI)~(\cmsq)] \leq 17.61$. From this range, a most probable value for the $\HI$ can be arrived at by considering the properties for the population of $\HI$ absorbers at low redshifts.

\begin{figure}
	\includegraphics[width=230pt,height=170pt,trim=0.1cm 0.40cm 0.4cm 0.5cm,clip=true]{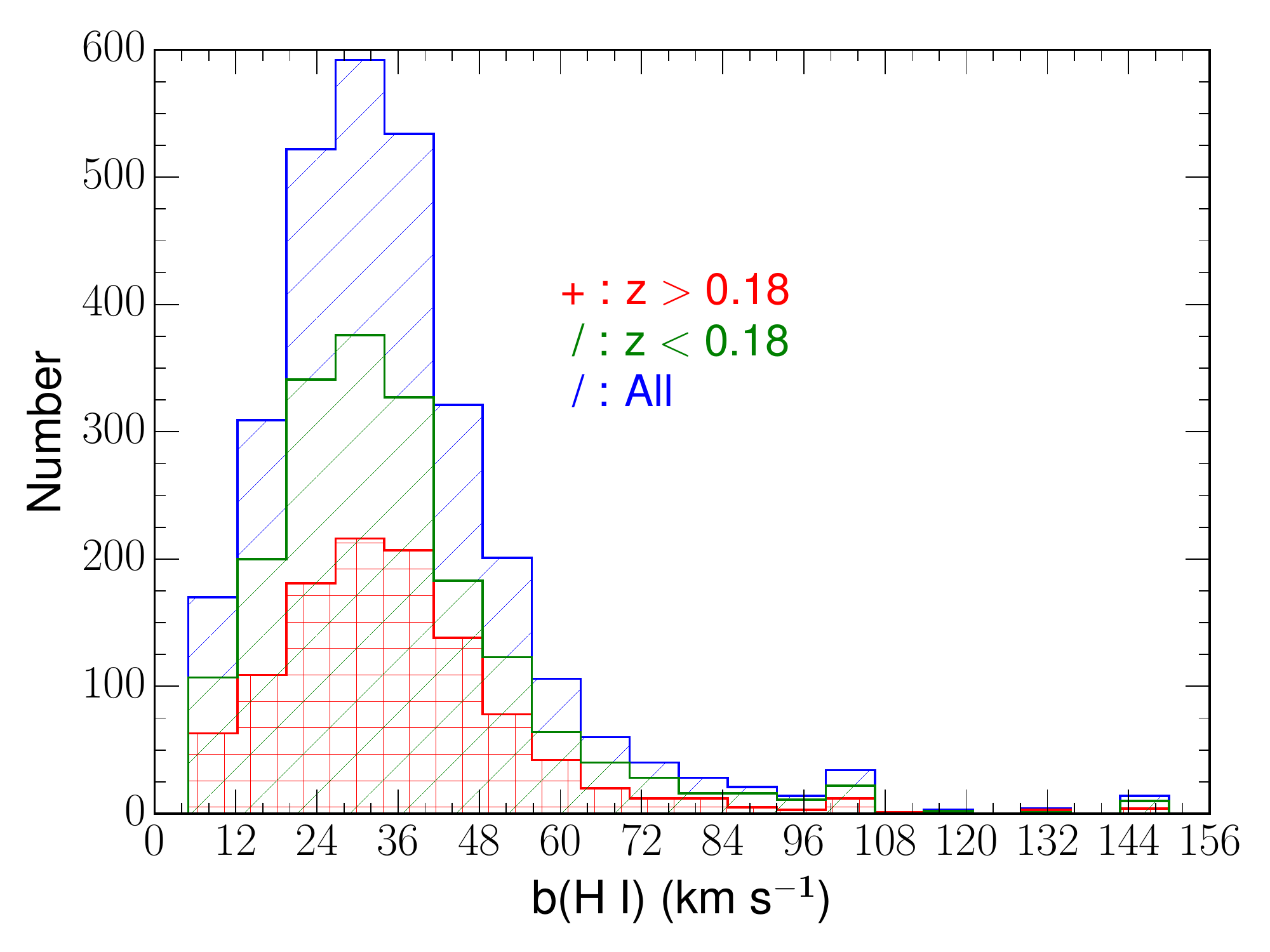}\quad
        \includegraphics[width=230pt,height=170pt,trim=0.2cm 0.46cm 0.4cm 0.47cm,clip=true]{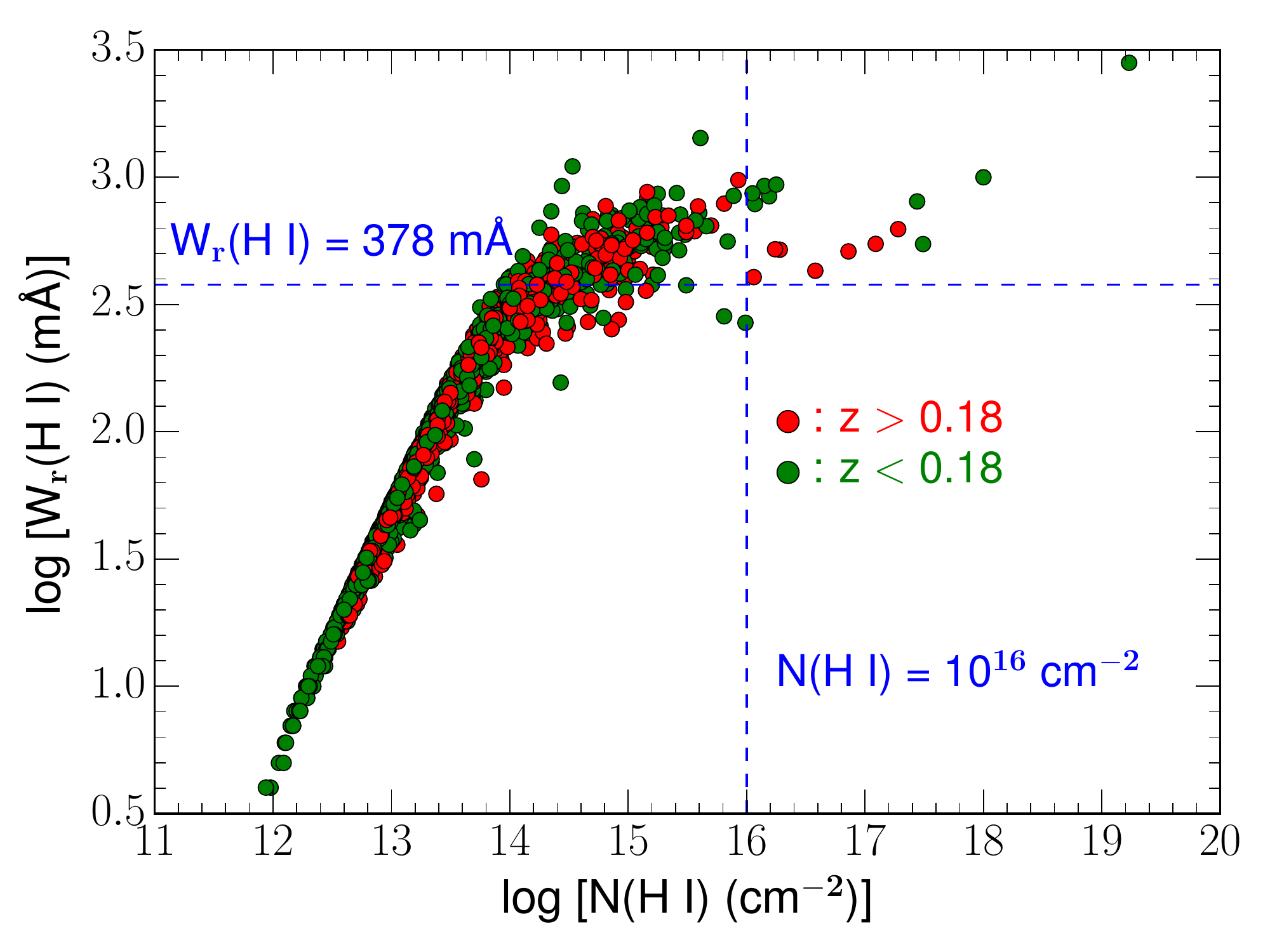}
    \caption{\textit{Top}: Distribution of Doppler $b$ parameters of $\Lya$ in the low redshift IGM survey by \citet{Danforth2016}. The histograms shaded in \textit{red} and \textit{green} represent the $\Lya$ at $z > 0.18$ and $z < 0.18$ respectively. The median value lies at $b(\HI) \sim 30~\kms$. The corresponding $N(\HI)$ value represents the most probable value for a low redshift $\Lya$ absorber. \textit{Bottom}: The $W_r(\HI)-N(\HI)$ plot for the sample in \citet{Danforth2016}. The \textit{red} and \textit{green} filled-circles represent the values for the $\Lya$ at $z > 0.18$ and $z < 0.18$ respectively. Amongst the lines with $W_r(\HI) > 378$~m{\AA}, only $\sim 6.3$\% have $\log~[N(\HI)~(\cmsq)] > 16.0$ for both the lower and higher redshift samples. This suggests that $\Lya$ in the local universe with such $W_r(\HI)$ values are unlikely to be strong.}
    \label{3}
\end{figure}

In the top panel of Figure~\ref{3}, we have compiled the $b$ measurements given by \citet{Danforth2016} for 2974 extragalactic $\Lya$ lines at $z_{abs} < 0.75$. For $z_{abs} > 0.18$, the coverage of other Lyman series lines allows a more robust estimate of $b$ and $N$ as compared to the $\Lya$ at lower redshifts. Since the systems in our study reside in the local universe, we have also looked at $b(\HI)$ distribution for $z_{abs} < 0.18$ and $z_{abs} > 0.18$ separately. The distribution of $b$-parameters has a median value of $b(\HI) \sim 30~\kms$ for the full sample as well as for the sub-sample at lower redshifts ($z_{abs} < 0.18$). This is also consistent with the $b(\HI)$ distributions in the STIS low redshift survey of $\Lya$ forest by \citet{Lehner2007} and that of CGM absorbers using COS data by \citet{Lehner2018} , where the median values for $b(\HI)$ are $\sim 30~\kms$ and $\sim 27~\kms$ respectively. Unresolved saturation affecting measurements of narrow and strong $\HI$ components is likely to be much less of an issue in the STIS sample because of its higher spectral resolution and cleaner line spread function compared to COS. The $b(\HI)$ distribution in Figure~\ref{3} suggests that there is only a $\sim 3$\% probability for the $b(\HI)$ to be as low as $10~\kms$. Similarly, we have also examined the $W_r(\HI)-N(\HI)$ relationship for the $\Lya$ in \citet{Danforth2016} at $z < 0.18$ and $z > 0.18$ which is shown in the bottom panel of Figure~\ref{3}. As we will see in the coming sections, all three of our systems have $W_r(\HI) \geq 378$~m{\AA}. Amongst such systems in \citet{Danforth2016}, only $\sim 6.3$\% are seen to have strong $\Lya$ ($\log~[N(\HI)~(\cmsq)] > 16.0$) for both low and high redshift samples. The core $N(\HI)$ in the absorber we are analyzing is likely to have its true column density nearer to the lower limit of $N(\HI) \sim 10^{14.6}~\cmsq$ corresponding to $b(\HI) \sim 35~\kms$.

\begin{figure*}
	\includegraphics[height=640pt,trim={1cm 1.3cm 3cm 5.5cm},clip=true]{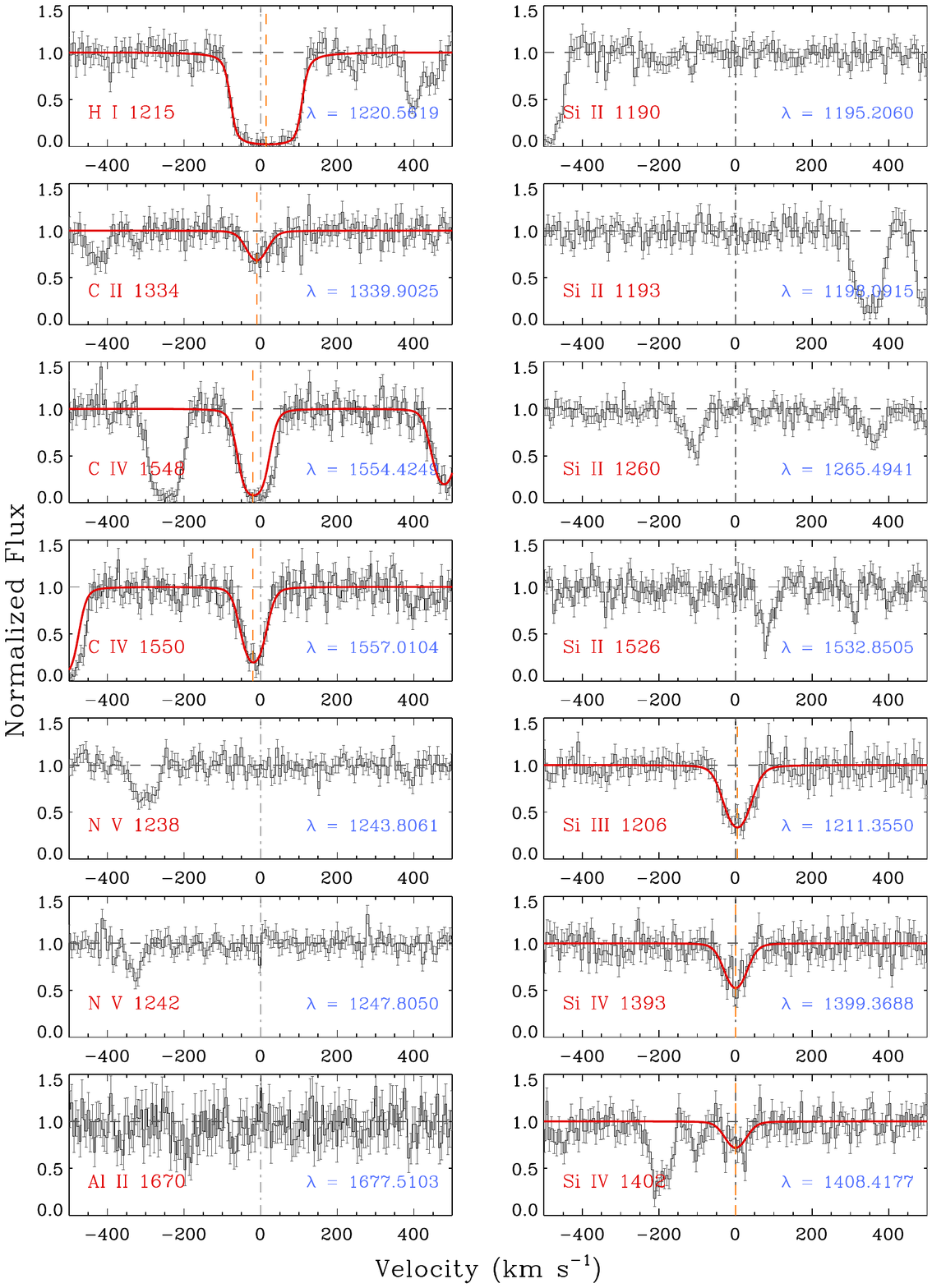}
    \caption{System plot of the $z_{abs}=0.00402$ $\CIV$ absorber towards SBS~$1122+594$ with free-fit of the saturated $\HI$. The zero velocity corresponds to the redshift of the absorber derived from the wavelength pixel that shows the maximum optical depth in the $\CIV$~$1548$ line represented by the \textit{dashed-dot} vertical line. The Y-axis is continuum normalized flux. The error bars represent $1\sigma$ uncertainty in flux values. The \textit{red} curves overplotted on top of the spectra represent the best-fit Voigt profiles and the output parameters are given in Table~\ref{tab2}. The \textit{dashed} vertical lines mark the line centroid given by the fitting routine.}
    \label{4}
\end{figure*}

\begin{table} 
\caption{Line measurements for the $z = 0.004024$ absorber towards SBS~$1122+594$.} 
\begin{center}
\begin{threeparttable}
\begin{tabular}{lccc}
\hline
Line     &	$W_r$   	   &       log~$[N_a~(\cmsq)]$	&	$[-v, +v]$     \\ 
         &      (m{\AA})           &                            &          ($\kms$)      \\   
\hline
$\HI$~$1215$		&   $819~{\pm}~74$      	&       $> 14.6$		& [-140, 160] \\
$\CII$~$1334$		&   $95~{\pm}~21$	&	$13.75~{\pm}~0.10$     	& [-110, 85] \\
$\AlII$~$1670$		&   $< 111$	        &	$< 12.5$ 	        & [-110, 85] \\
$\SiII$~$1190$          &   $< 57$		&	$< 13.2$	        & [-110, 85] \\
$\SiII$~$1193$          &   $< 57$		&	$< 12.9$	        & [-110, 85] \\
$\SiII$~$1260$          &   $< 39$		&	$< 12.4$	        & [-55, 85] \\
$\SiII$~$1526$          &   $< 60$		&	$< 13.4$	        & [-110, 35] \\
$\CIV$~$1548$           &   $> 562$		&	$> 14.51$		& [-110, 85] \\
$\CIV$~$1550$ 	        &   $434~{\pm}~20$     &	$14.56~{\pm}~0.08$      & [-110, 85] \\
$\NV$~$1238$	        &   $< 45$             &  	$< 13.3$	        & [-110, 85] \\
$\NV$~$1242$            &   $< 45$             &	$< 13.7$	        & [-110, 85] \\
$\SiIII$~$1206$		&   $268~{\pm}~17$	& 	$13.29~{\pm}~0.07$	& [-110, 85] \\	
$\SiIV$~$1393$		&   $160~{\pm}~22$	&	$13.37~{\pm}~0.08$	& [-65, 55] \\
$\SiIV$~$1402$		&   $81~{\pm}~65$	&	$13.40~{\pm}~0.77$	& [-65, 55] \\
Line     &	$v$      &       log~$[N~(\cmsq)]$	&    $b$     \\
         &     ($\kms$)    &                              &  ($\kms$)    \\   
$\HI$~$1215$	        &   $14~{\pm}~3$	&	$(15.21 - 17.25)$	 &	$(31 - 50)$	\\
$\CII$~$1334$	        &   $-10~{\pm}~6$	&	$13.77~{\pm}~0.08$	 &	$32~{\pm}~9$	\\
$\CIV$~$1548-1550$	&   $-20~{\pm}~2$	&	$14.49~{\pm}~0.04$	 &	$31~{\pm}~3$	\\
$\SiIII$~$1206$         &   $4~{\pm}~4$        &	$13.25~{\pm}~0.05$	 &	$36~{\pm}~5$	\\
$\SiIV$~$1393-1402$     &   $0~{\pm}~6$        &	$13.40~{\pm}~0.10$	 &	$33~{\pm}~10$	\\
\hline
\end{tabular}
\label{tab2}
\begin{tablenotes}
\item[] The upper part of the table presents the apparent optical depth measurements for the various lines in the rest-frame of the absorber and the lower part consists of the Voigt fitting parameters. The $\CIV$~$1548$ suffers from contamination for the part of the profile with $v > 0$.
\end{tablenotes}
\end{threeparttable}
\end{center}
\end{table}

\subsection{The $\mathbf{z_{abs}=0.00402}$ absorber towards SBS~$\mathbf{1122+594}$}

The absorber is detected in $\HI$, $\CII$, $\CIV$, $\SiIII$ and $\SiIV$ at $\geq 3\sigma$ whereas $\AlII$, $\SiII$ and $\NV$ are non-detections (see Table~\ref{tab2}). The $N_a(v)$ comparison of Figure~\ref{2} for the $\CIVdblt$ lines indicate contamination in the velocity interval $+5 \lesssim \Delta v \lesssim +45~\kms$ in the $\CIV$~$1548$ line. While performing simultaneous profile fitting on the $\CIV$ lines, we deweight these contaminated pixels to exclude them from the fitting procedure. The $N_a(v)$ comparison (Figure~\ref{2}) also shows mild saturation in the $\CIV$~$1548$ line core which the simultaneous profile fit takes into account. The resultant fit model is shown in Figure~\ref{4}. The metal lines do not show any evidence for significant sub-component structure. The model fits were therefore generated using a single component. Similar line widths for the metal lines indicate turbulence dominating the line broadening ($b_{nt}/b \gtrsim 89$\%), with $T \leq 7 \times 10^{5}$~K. A single component model also fits the broad and saturated $\Lya$, though the fit is not exclusive because of strong line saturation. As done for the previous absorber, the $b(\HI)$ values were allowed to vary between pure thermal and pure non-thermal broadening scenarios using the $b(\CIV)$ value as reference. It was found that $b(\HI) > 50~\kms$ is too broad to fit the observed $\Lya$. The admissible $b$-values fall in the range $b(\HI) = 31 - 50~\kms$ with a corresponding wide column density range of $\HI$ as $15.21 \leq \log~[N(\HI)~(\cmsq)] \leq 17.25$. The most probable $b(\HI)$ value of $\sim 30~\kms$ given by the large sample of low-$z$ $\HI$ absorbers (see Figure~\ref{3} and Sec.~\ref{Danforth}) suggests $\log~[N(\HI)~(\cmsq)] \lesssim 17.25$. In addition, the rest-frame equivalent width of $W_r(\CII~1335) = 95~{\pm}~21$~m{\AA} (see Figure~\ref{2}) makes this a weak $\MgII$ class of absorber which are associated with sub-Lyman limit systems \citep{Churchill1999,Rigby2002,Narayanan2005,Muzahid2018}. Considering both these, the true column density is presumably closer to, but lower than $\log~[N(\HI)~(\cmsq)] \leq 17.25$.

\begin{table} 
\caption{Line measurements for the $z = 0.00574$ absorber towards RXJ~$1230.8+0115$.} 
\begin{center}
\begin{threeparttable}
\begin{tabular}{lccc}
\hline
Line     &	$W_r$   	   &       log~$[N_a~(\cmsq)]$	&	$[-v, +v]$     \\ 
         &      (m{\AA})           &                            &          ($\kms$)      \\   
\hline
$\HI$~$1215$		&   $616~{\pm}~12$      &	$> 14.5$	& [-140, 90] \\
$\CII$~$1334$		&   $81~{\pm}~8$	&	$13.67~{\pm}~0.08$      & [-70, 40] \\
$\AlII$~$1670$		&   $< 82$	&	$< 12.3$ 	& [-70, 40] \\
$\SiII$~$1190$          &   $< 32$		&	$<13.0$			& [-70, 40] \\
$\SiII$~$1260$          &   $57~{\pm}~9$        &	$12.58~{\pm}~0.07$	& [-70, 40] \\
$\CIV$~$1548$           &   $67~{\pm}~9$	&	$13.27~{\pm}~0.09$	& [-70, 40] \\
$\CIV$~$1550$ 	        &   $57~{\pm}~9$      	&	$13.48~{\pm}~0.09$      & [-70, 40] \\
$\NV$~$1238$	        &   $< 12$             &  	$< 12.7$	        & [-70, 40] \\
$\NV$~$1242$            &   $< 12$             &	$< 12.9$	        & [-70, 40] \\
$\SiIII$~$1206$		&   $198~{\pm}~11$	& 	$13.14~{\pm}~0.07$	& [-70, 40] \\	
$\SiIV$~$1393$		&   $71~{\pm}~8$	&	$12.96~{\pm}~0.07$	& [-70, 40] \\
$\SiIV$~$1402$		&   $34~{\pm}~8$	&	$12.90~{\pm}~0.08$	& [-70, 40] \\
Line     &	$v$      &       log~$[N~(\cmsq)]$	&    $b$     \\
         &     ($\kms$)    &                              &  ($\kms$)    \\   
$\HI$~$1215$	        &   $-29$	        &	$(14.88 - 16.92)$	 &	$(18 - 30)$	\\
                        &   $6$	                &	$(14.62 - 15.24)$	 &	$(23 - 31)$	\\
$\CII$~$1334$	        &   $-28~{\pm}~3$	&	$13.15~{\pm}~0.06$	 &	$8~{\pm}~5$	\\
                        &   $6~{\pm}~2$	        &	$13.64~{\pm}~0.03$	 &	$9~{\pm}~3$	\\
$\CIV$~$1548-1550$      &   $-35~{\pm}~4$	&	$12.51~{\pm}~0.10$	 &	$9*$	\\
                        &   $-2~{\pm}~1$	&	$13.16~{\pm}~0.06$	 &	$9~{\pm}~3$	\\
$\SiII$~$1260$          &   $-35~{\pm}~4$       &	$11.91~{\pm}~0.12$	 &	$11~{\pm}~9$	\\
                        &   $3~{\pm}~2$         &	$12.52~{\pm}~0.04$	 &	$9~{\pm}~3$	\\
$\SiIII$~$1206$         &   $-29~{\pm}~3$       &	$12.61~{\pm}~0.05$	 &	$11~{\pm}~7$	\\
                        &   $6~{\pm}~2$         &	$13.25~{\pm}~0.07$	 &	$9~{\pm}~4$	\\
$\SiIV$~$1393-1402$     &   $-32~{\pm}~3$       &	$12.27~{\pm}~0.09$	 &	$11*$	\\
                        &   $0~{\pm}~2$         &	$12.85~{\pm}~0.03$	 &	$10~{\pm}~3$	\\
\hline
\end{tabular}
\label{tab3}
\begin{tablenotes}
\item[] The upper part of the table has the apparent optical depth measurements for the various lines in the rest-frame of the absorber and the lower part consists of the Voigt fitting parameters. The $\SiII$~$1193$ line is not included in the table as its wavelength region is strongly contaminated by Galactic ISM $\NI$~$1200$ features. The $b(\CIV)$ and $b(\SiIV)$ for Cloud 1 are treated as fixed parameters and this is represented with $*$.
\end{tablenotes}
\end{threeparttable}
\end{center}
\end{table}

\begin{figure*}
	\includegraphics[height=640pt,trim={1cm 1.3cm 3cm 5.5cm},clip=true]{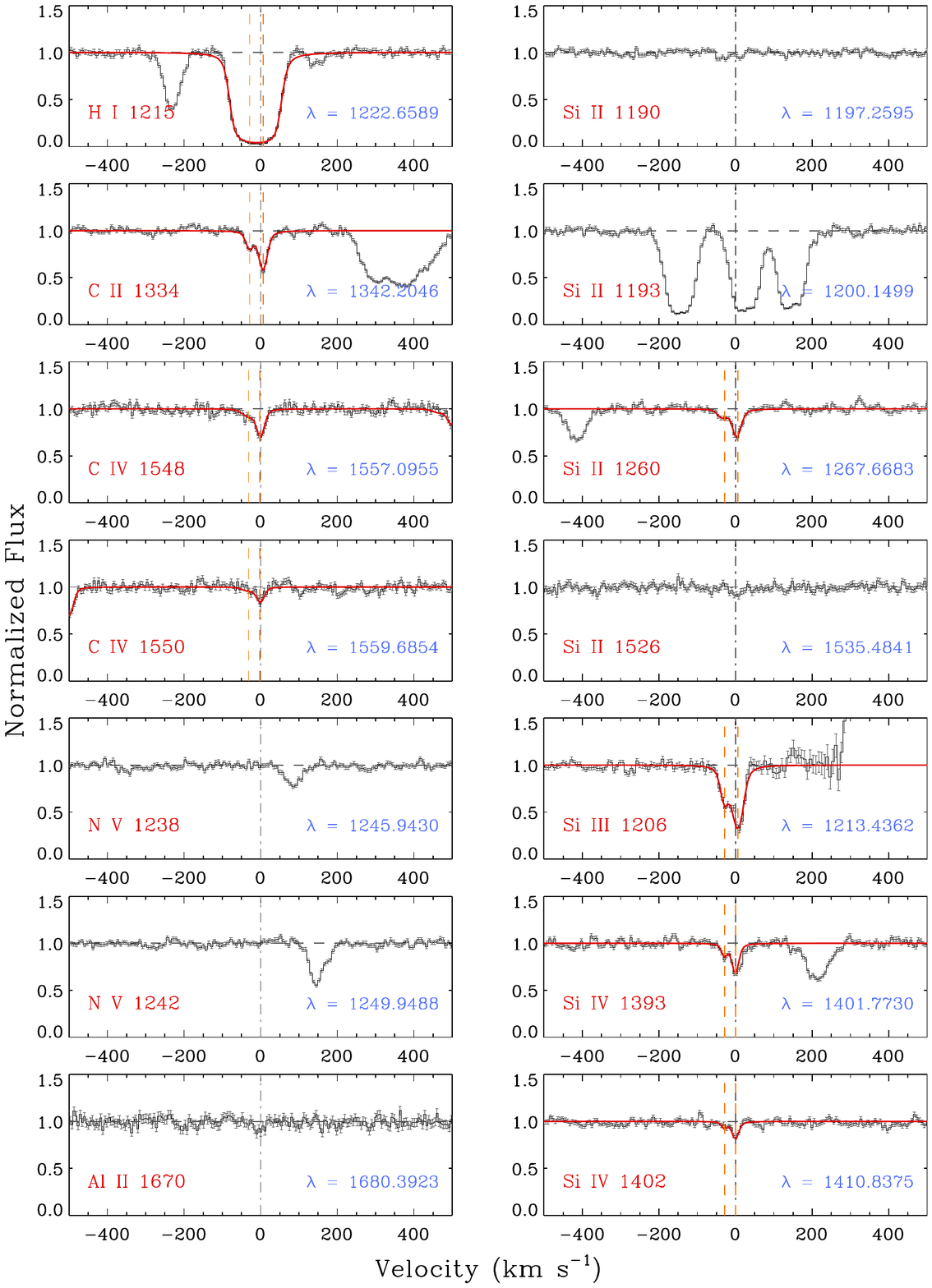}
    \caption{System plot of the $z_{abs}=0.00574$ $\CIV$ absorber towards RXJ~$1230.8+0115$ with free-fit of the saturated $\HI$. The zero velocity corresponds to redshift of the absorber derived from the wavelength pixel that shows maximum optical depth in the $\CIV$~$1548$ line, represented by the \textit{dashed-dot} vertical line. The Y-axis is continuum normalized flux. The error bars represent $1\sigma$ uncertainty in flux values. The overplotted \textit{red} curves represent the Voigt profile fits. The fit parameters are given in Table~\ref{tab3}. The \textit{dashed} vertical lines mark the centroid of the two line components obtained from fitting.}
    \label{5}
\end{figure*}

\subsection{The $\mathbf{z_{abs}=0.00574}$ absorber towards RXJ~$\mathbf{1230.8+0115}$}

The absorber is detected in $\HI$, $\CII$, $\CIV$, $\SiII$, $\SiIII$ and $\SiIV$ at $\geq 3\sigma$ (See Table~\ref{tab3}). The metal lines show two kinematically distinct components, which are evident in the apparent column density comparison plots of Figure~\ref{2} and the system plot shown in Figure~\ref{5}. We refer to these separate components as Cloud 1 and Cloud 2 at $v(\CIV) = -34~\kms$ and $v(\CIV) = -2~\kms$ respectively, which are the velocities in the rest frame of the absorber derived from free-fits to the metal lines. The saturated $\Lya$ line is modelled by fixing the velocities of the components to those of $\CII$~$1335$ since there is no unique solution for $\Lya$ that can be arrived at through a free-fit. For Cloud 1, the $b(\CIV)$ implies $T \leq 2 \times 10^{5}$~K. The possible range for $b(\HI)$ allowed by the metal lines is $(9 - 31)~\kms$ where the limits are from assuming pure non-thermal and thermal broadening scenarios respectively. However, Voigt profile models synthesized with $b(\HI) < 23~\kms$ are too narrow for a good fit, which narrows the possible range of $b(\HI)$ to $23 - 31~\kms$. The corresponding column density range is $14.62 \leq \log~[N(\HI)~(\cmsq)] \leq 15.24$. Similarly, for Cloud 2 we obtain the upper limit for the cloud temperature as $T \leq 6 \times 10^{4}$~K and $b(\HI)$ as $18 - 62~\kms$. However, $b(\HI) > 30~\kms$ are too broad to fit the data. Thus, the $b(\HI)$ in Cloud 2 can vary from $18~\kms$ to $30~\kms$ with a corresponding column density range of $14.88 \leq \log~[N(\HI)~(\cmsq)] \leq 16.92$. Within this range, the $N(\HI)$ is likely to be closer to the lower limit if we choose the most probable $b(\HI) \sim 30~\kms$ as explained in Sec.~\ref{Danforth}. \citet{Rosenberg2003} have also measured the metal lines and have derived a range for the $\HI$ column densities in the two components using $HST$/STIS and $FUSE$ data with access to some of the higher order Lyman series lines. The two component profile is clearly evident in the metal lines as seen by the higher resolution of STIS, with the narrow line widths consistent with photoionized gas. The $\log~N(\HI) = 15 - 17.8$ determined by \citet{Rosenberg2003} through measurements of $FUSE$ higher order Lyman lines is consistent with the range that we obtain for the $\HI$ column densities. The quality of the $FUSE$ spectrum was inadequate to make an exact estimate on $N(\HI)$ in the two components. The $\log~N(\HI) = 16.2$ which \citet{Rosenberg2003} adopt for modelling Cloud 1 is consistent with the range that we have arrived at. For Cloud 2, their adopted value is a factor of 25 more than the upper limit on $\HI$ that we obtain. This difference possibly stems from the fact that their profile fits to $\HI$ are based on an assumed metallicity of $-1.2$~dex for either clouds, and only approximate because of the quality of the $FUSE$ data. The densities they arrive at from ionization modelling are nonetheless comparable to the range that we obtain from the models.

\section{DENSITY AND TEMPERATURE FROM IONIZATION MODELLING} \label{Sec4}

We performed photoionization modelling on the absorbers using \textsc{CLOUDY} v13.03 \citep{Ferland2013}. The $\HI$ column density in all the three absorbers carry a significant uncertainty because of saturation in $\Lya$, the only $\HI$ transition covered by the archival COS observations. This rules out accurate metallicity estimations based on the models. However,\textsc{CLOUDY} provides useful constraints on gas phase density and photoionization equilibrium temperatures, which can be compared for consistency with temperatures provided by the line widths.\textsc{CLOUDY} models assume the absorbing gas cloud to be static (no expansion), isothermal, with a plane parallel geometry, and no dust content. The model cloud is assumed to be photoionized by the extragalactic UV background (EUB) light at the redshift of these absorbers. We used the EUB model given by \citet{Khaire2018} (fiducial Q18model; hereafter KS18), instead of the earlier \citet{Haardt2012} models. The former incorporates updated values of cosmic star formation rate density and far-UV extinction from dust \citep{Khaire2015a}, along with most recent estimates of emissivity of QSOs \citep{Khaire2015b}, and the distribution of $\HI$ in the IGM \citep{Inoue2014}. As opposed to the Haardt \& Madau 2012 background, the KS18 model is consistent with the recent $z < 0.5$ photoionization rate measurements of \citet{Shull2015} and \citet{Gaikwad2016}. In the photoionization models, the relative abundances of heavy elements are initially assumed to be solar as given by \citet{Asplund2009}.\textsc{CLOUDY} models were run in each case for the respective upper and lower limits of $\HI$ column densities. A suite of ionization models were generated for metallicities from [X/H] = $-6.0$ to [X/H] = $2.0$ in steps of 0.1 dex, for densities ranging from $10^{-6}~\cc \leq n_{\mathrm{H}} \leq 10^{-1}~\cc$.

\subsection{Densities and temperatures for the $\mathbf{z_{abs}=0.00346}$ absorber towards PG~$\mathbf{1148+549}$}
\begin{figure*}
	    \includegraphics[width=206pt,trim=3cm 2.76cm 2.1cm 7.8cm,clip=true,angle=90]{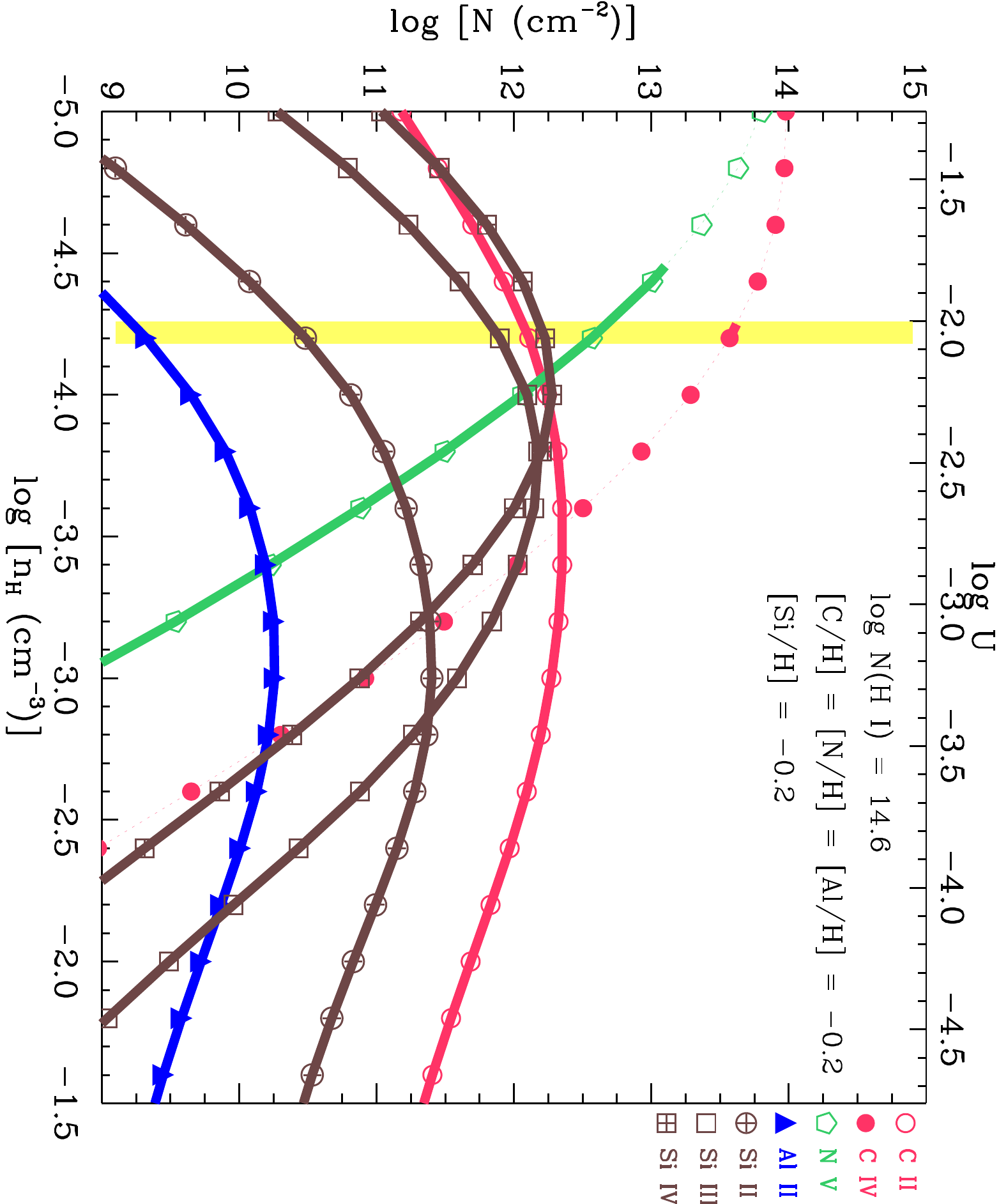}\quad
        \includegraphics[width=206pt,trim=3cm 2.92cm 2.1cm 7.8cm,clip=true,angle=90]{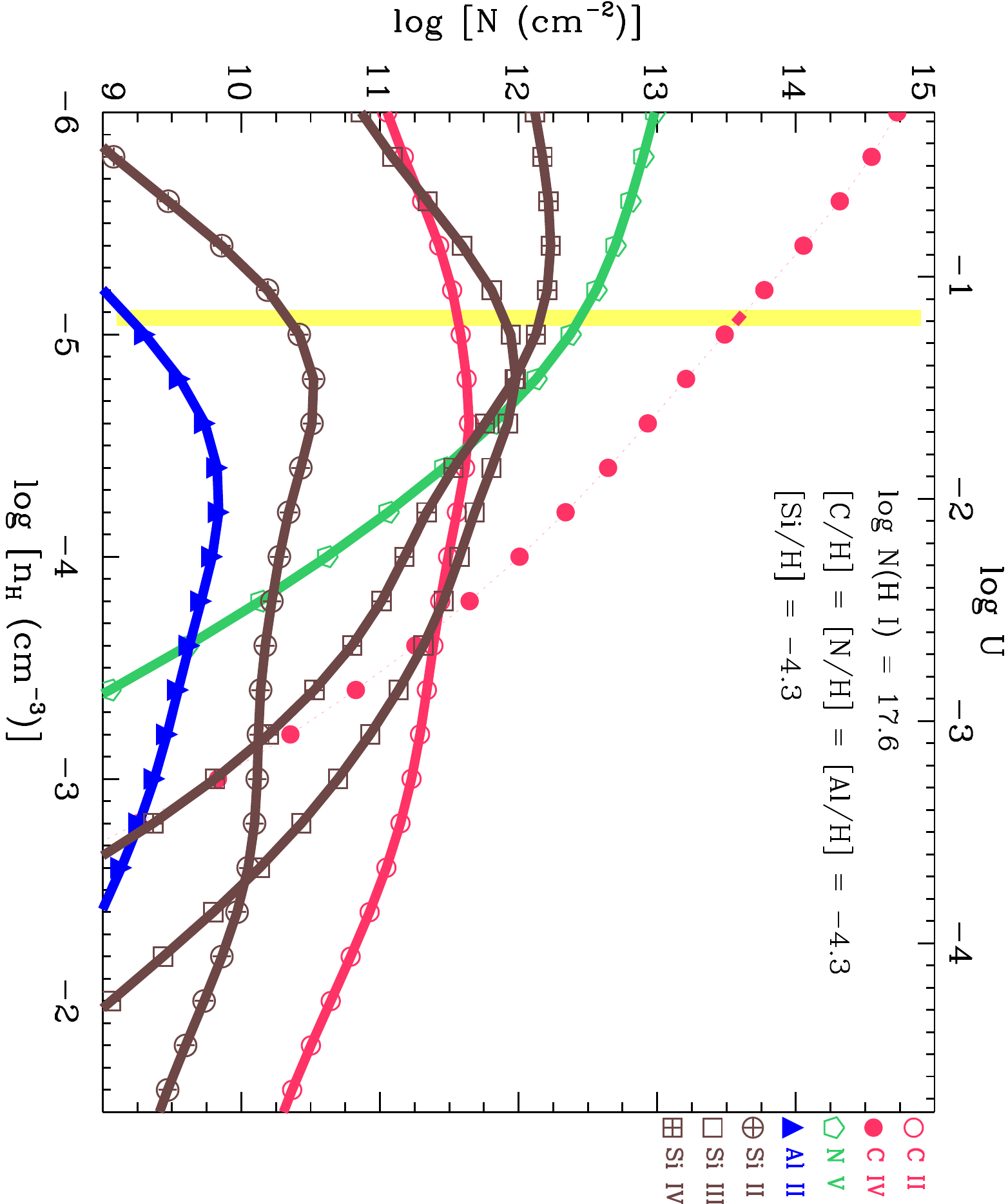}
    \caption{The photoionization equilibrium models for $z_{abs}=0.00346$ towards PG~$1148+549$ for $N(\HI) = 14.6, 17.6$~dex. The vertical axis corresponds to logarithm of column densities for the various ions as predicted by the models. The horizontal axis represents density ranging from log~$[n_{H}~(\cc)] = -5$ to log~$[n_{H}~(\cc)] = -1$ for the plot on the \textit{left} and from log~$[n_{H}~(\cc)] = -6$ to log~$[n_{H}~(\cc)] = -1$ for the plot on the \textit{right}. The model predictions for the different ions are plotted with different symbols and the \textit{thin} curve joining them. The \textit{thick} portion of the curves indicate the $1\sigma$ range of the observed column density for the respective ions. The yellow strip highlights the narrow range of densities for which the models are consistent with the observed $N(\CIV)$ and the upper limits derived from the non-detections of the other species. The [C/H] in the absorber is constrained from $N(\CIV)$.} 
    \label{6}
\end{figure*}

The $\HI$ column density in this absorber falls within the wide range of $14.60 \leq \log~[N(\HI)~(\cmsq)] \leq 17.61$. The photoionization equilibrium models for the lower limit on $\HI$ column density of $\log~[N(\HI)~(\cmsq)]= 14.6$ is shown in Figure~\ref{6}. Assuming the [C/Si] abundance to be solar, the observed $\log~[N(\CIV)/N(\SiIV)] \gtrsim 1.2$ is valid only for densities of $n_{\mathrm{H}} < 6.3 \times 10^{-5}~\cc$ (see Figure~\ref{6}). This upper limit is close to the density where the ionization fraction of $\SiIV$ peaks. At this density, the [C/H] $= -0.2$ for the model's prediction to match the observed $N(\CIV)$. At lower densities, the [C/H] $< -0.2$. Thus, the carbon and silicon abundance in this absorber is constrained to $\leq -0.2$~dex, assuming solar relative elemental abundance pattern. At the limiting density of $n_{\mathrm{H}} = 6.3 \times 10^{-5}~\cc$, the single phase model predicts an equilibrium temperature of $T = 1.6 \times 10^{4}$~K, $p/k = 1.1$~K~$\cc$, total hydrogen column density of $\log~[N(\mathrm{H})~(\cmsq)] = 17.9$, and a line of sight thickness of $L = 4$~kpc. The photoionization temperature from the models agrees with the upper limit of $7 \times 10^{4}$~K given by the $\CIV$ $b$-parameter.

The models based on the upper limit on the $\HI$ column density of $\log~[N(\HI)~(\cmsq)] = 17.6$ is shown in Figure~\ref{6}. The observed $\log~[N(\CIV)/N(\SiIV)] \gtrsim 1.2$ is valid for $n_{\mathrm{H}} < 0.8 \times 10^{-5}~\cc$. This limits the carbon and silicon abundance to [C/H] = [Si/H] = $\lesssim -4.3$, for a [C/Si] of solar. The models for $n_{\mathrm{H}} \leq 0.8 \times 10^{-5}~\cc$ also predict exceedingly high path lengths of $L > 570$~Mpc, which are physically unrealistic for an absorber with very little kinematic complexity. The assumed high $\HI$ column density is what brings about the large path length for this absorber, which implies that the true $\HI$ column density is significantly lower than this. It is more likely that the true column density is closer to the $\HI$ lower limit of $14.6$~dex, as indicated in Sec.~\ref{Danforth}. The ionization modelling is thus able to suggest a narrow range for the physical properties and abundances in this absorber (see Table~\ref{tab4}), despite the uncertainty in $\HI$ due to line saturation.

\begin{figure*}
	    \includegraphics[width=206pt,trim=3cm 2.77cm 2cm 7.83cm,clip=true,angle=90]{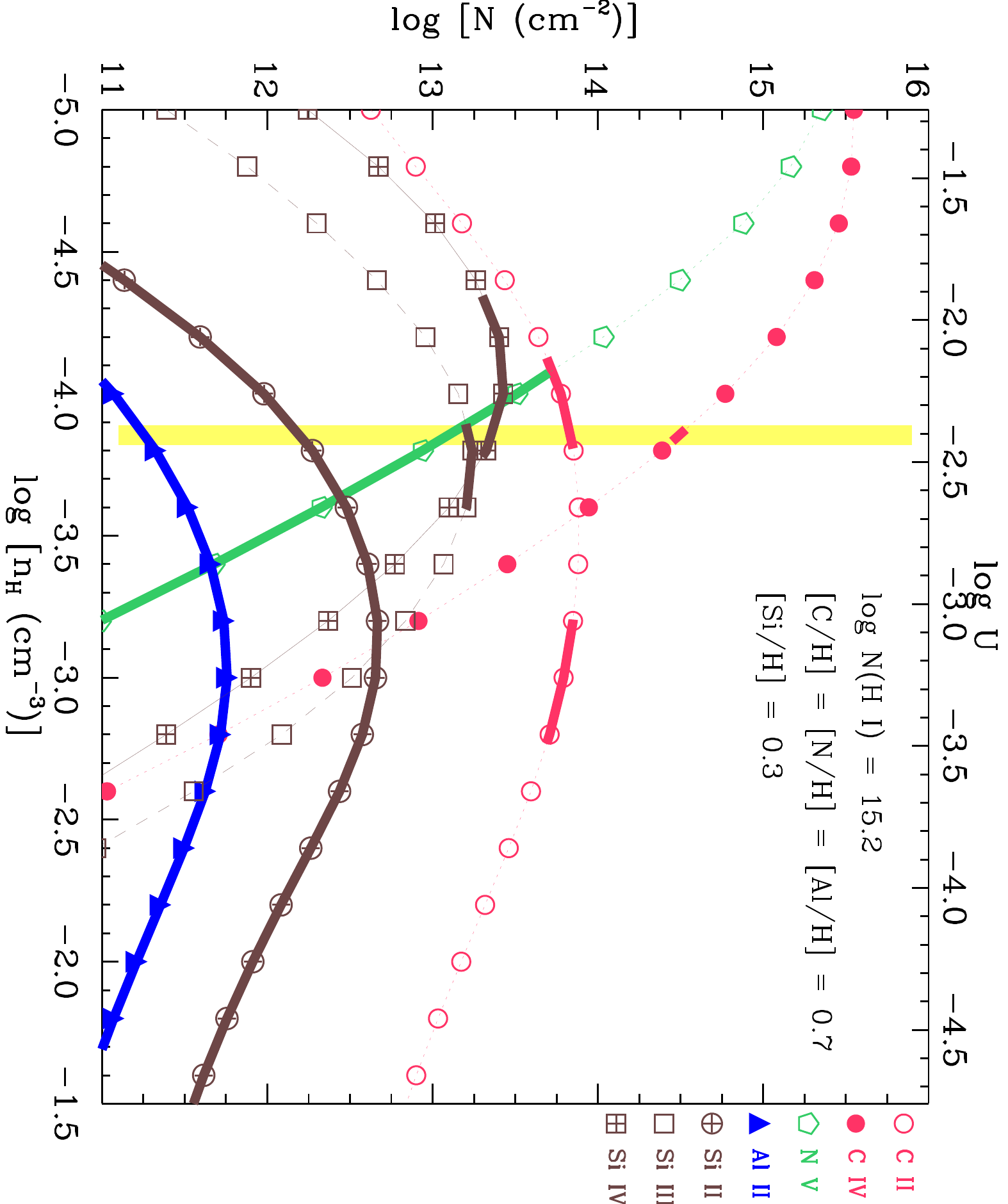} \quad
        \includegraphics[width=206pt,trim=3cm 2.77cm 2cm 7.83cm,clip=true,angle=90]{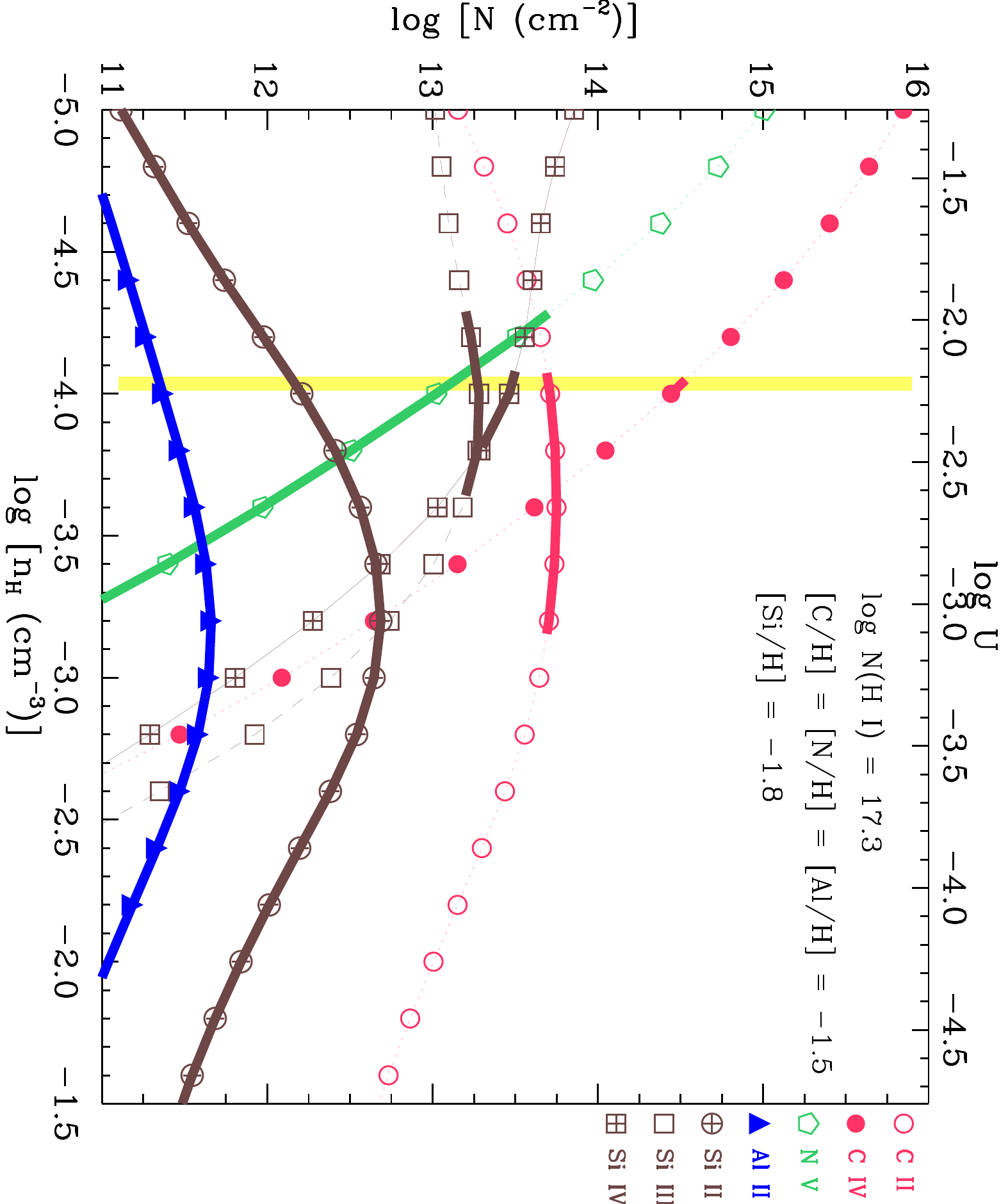}
    \caption{The photoionization equilibrium models for $z_{abs}=0.00402$ towards SBS~$1122+594$ for $N(\HI) = 15.2, 17.3$~dex. The vertical axis corresponds to logarithm of column densities for the various ions as predicted by the models. The horizontal axis represents density ranging from log~$[n_{H}~(\cc)] = -5$ to log~$[n_{H}~(\cc)] = -1$. The model predictions for the different ions are plotted with different symbols and the \textit{thin} curve joining them. The \textit{thick} portion of the curves indicate the $1\sigma$ range of the observed column density for the respective ions. The yellow strip highlights the narrow range of densities for which the models are consistent with the observed $N(\CIV)$, $N(\CII)$, $N(\SiIII)$, and $N(\SiIV)$, along with upper limits from the non-detections of the other metal species.}
    \label{7}
\end{figure*}

\begin{figure*}
	    \includegraphics[width=206pt,trim=3cm 2.77cm 2cm 7.83cm,clip=true,angle=90]{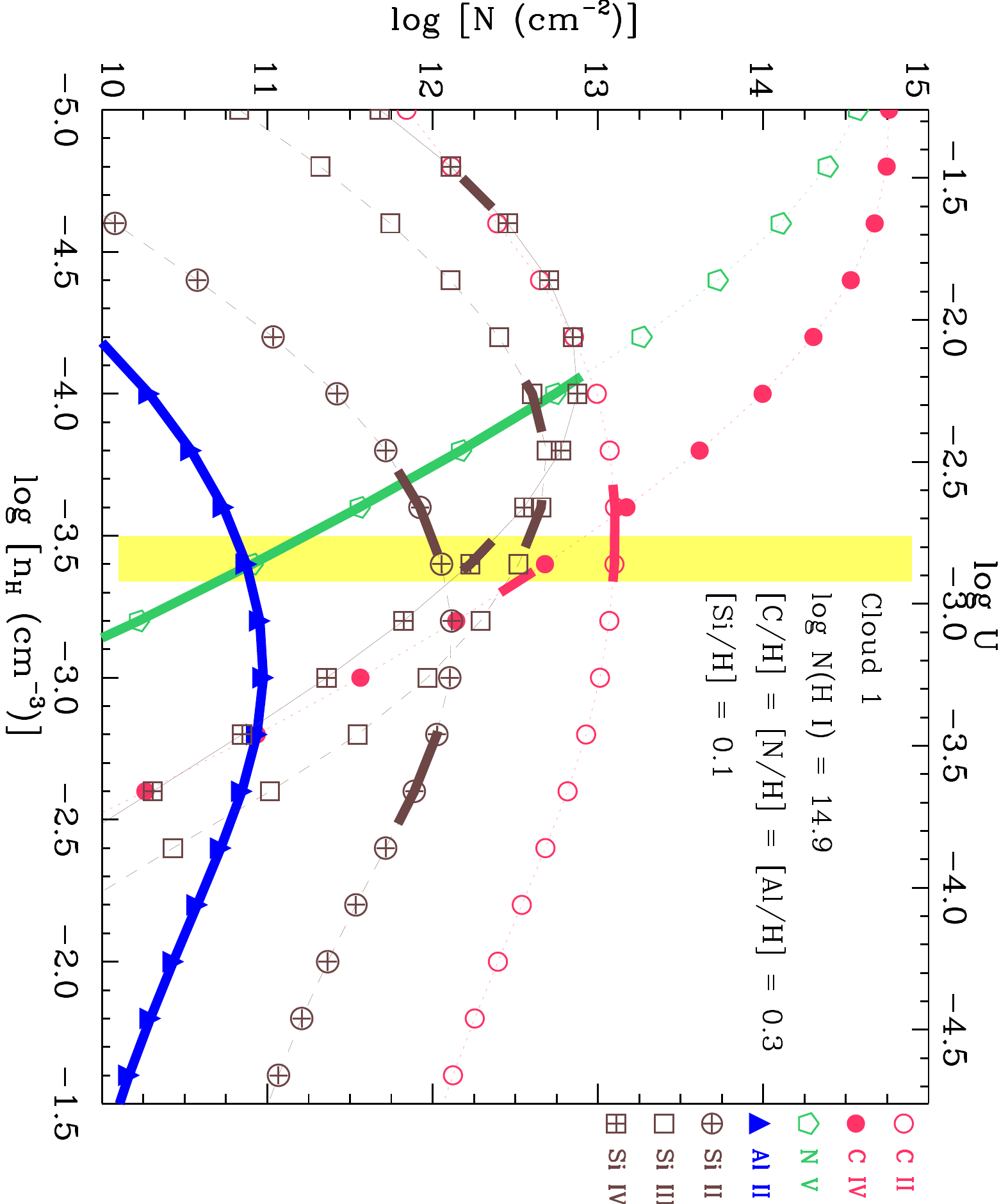} \quad
        \includegraphics[width=206pt,trim=3cm 2.77cm 2cm 7.83cm,clip=true,angle=90]{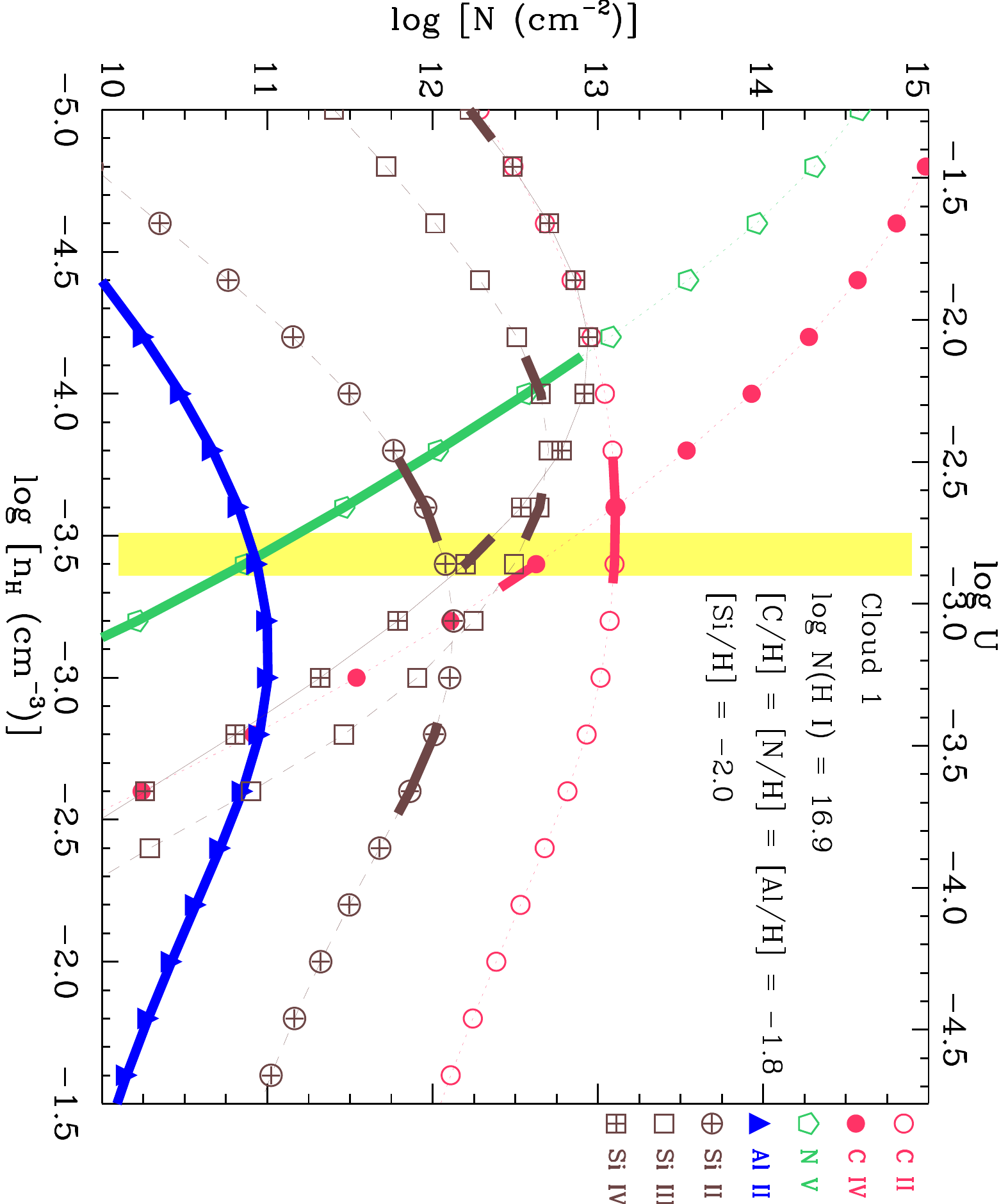}
        \includegraphics[width=206pt,trim=3cm 2.77cm 2cm 7.83cm,clip=true,angle=90]{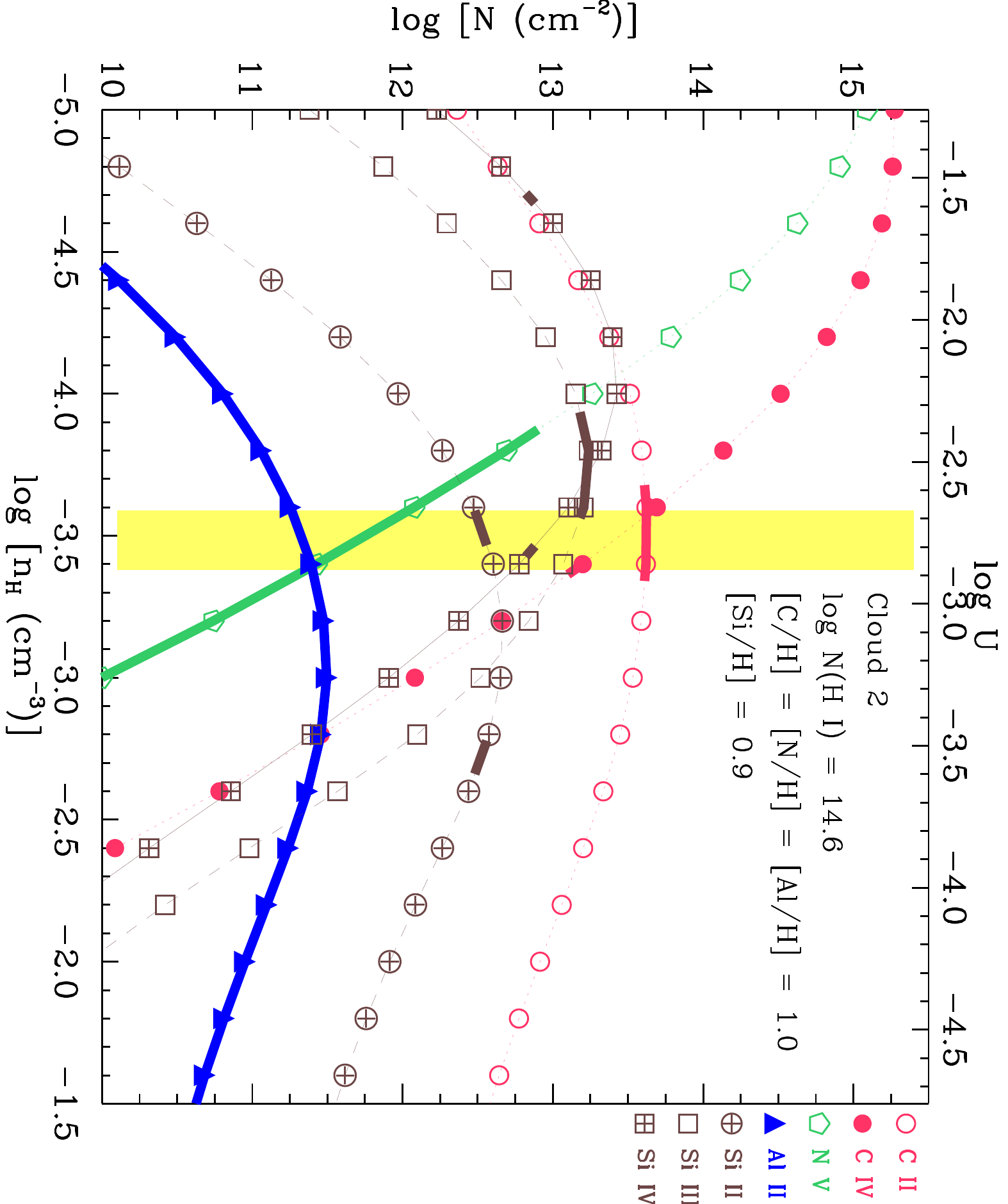} \quad
        \includegraphics[width=206pt,trim=3cm 2.77cm 2cm 7.83cm,clip=true,angle=90]{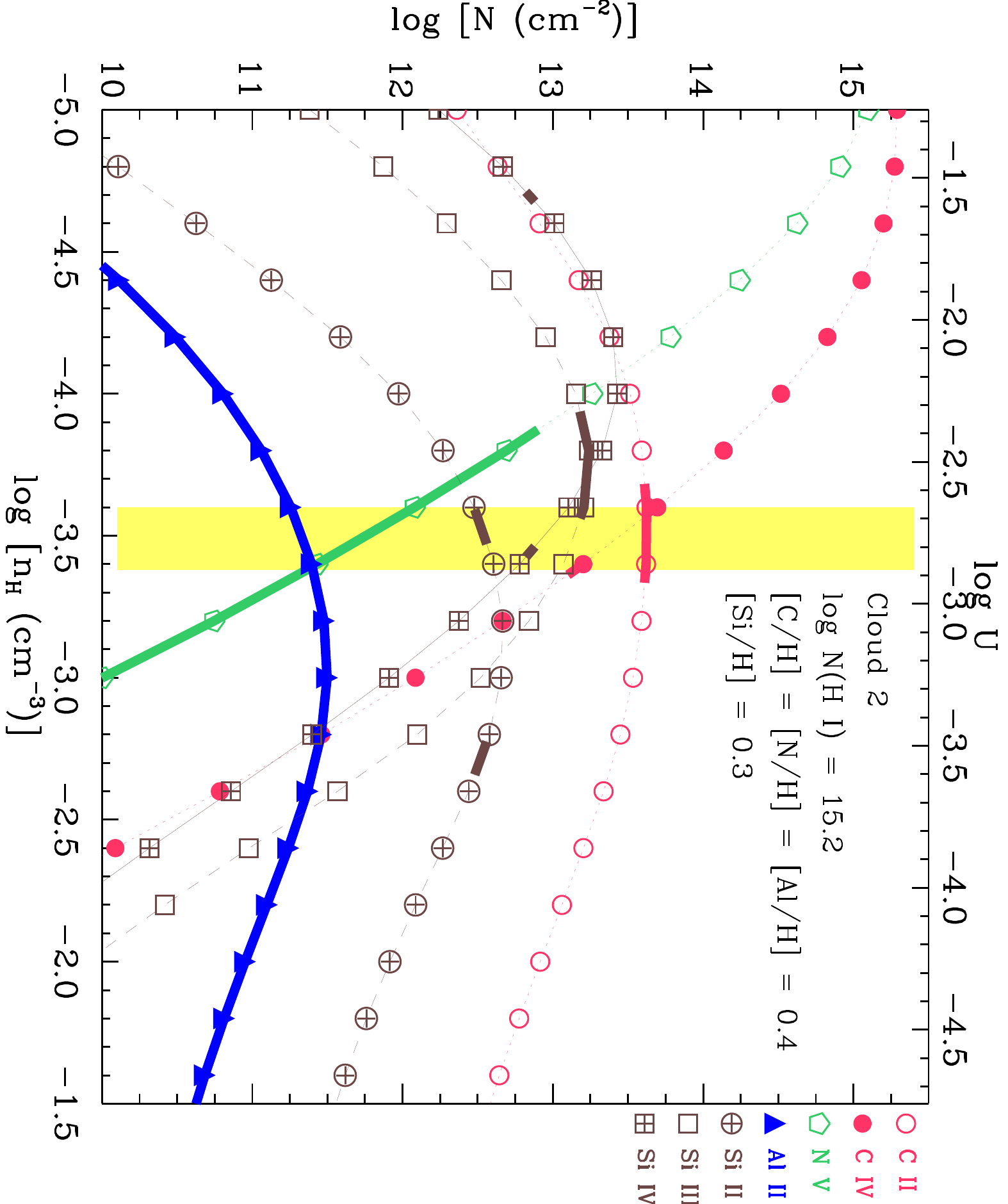}
    \caption{The photoionization equilibrium models for the clouds in the $z_{abs}=0.00574$ absorber towards RXJ~$1230.8+0115$ for $N(\HI) = 14.9, 16.9$~dex and $N(\HI) = 14.6, 15.2$~dex in Cloud 1 and Cloud 2 respectively. The vertical axis corresponds to logarithm of column densities for the various ions as predicted by the models. The horizontal axis represents density ranging from log~$[n_{H}~(\cc)] = -5$ to log~$[n_{H}~(\cc)] = -1$. The model predictions for the different ions are plotted with different symbols and the \textit{thin} curve joining them. The \textit{thick} portion of the curves indicate the $1\sigma$ range of the observed column density for the respective ions. The yellow strip highlights the range of densities for which the models are consistent with the observed column densities for the detected species and the upper limits from the non-detections of other ions.}
    \label{8}
\end{figure*}

\begin{table*} 
\begin{center}
\caption{Summary of the results from photoionization modelling.}
\begin{threeparttable}
\begin{tabular}{lcccccc}
\hline
QSO      &	$z_{abs}$      &       log~$[N(\HI)~(\cmsq)]$	& log~$[N(\mathrm{H})~(\cmsq)]$  &	[C/H]       &      $n_{\mathrm{H}}~(\cc)$      &     $T$~(K) \\   
\hline
PG~$1148+549$   &    $0.003$   &         $(14.6 - 17.6)$  &  $(17.9 - 22.2)$  &  $< -0.2$   &    $(0.8 - 6.3) \times 10^{-5}$  &  $(1.6 - 4.2) \times 10^{4}$\\
SBS~$1122+594$  &    $0.004$   &         $(15.2 - 17.3)$  &  $(18.1 - 20.6)$  &  $(-1.5 - +0.7)$    &    $(0.9 - 1.5) \times 10^{-4}$ &   $(1.3 - 2.4) \times 10^4$\\
RXJ~$1230.8+0115$  &   $0.005$    &      $(14.9 - 16.9)$  &  $(17.3 - 19.6)$  &  $(-1.8 - +0.3)$    &    $(3.1 - 4.6) \times 10^{-4}$ &  $(1.4 - 1.9) \times 10^{4}$\\  
                   &   $0.005$    &      $(14.6 - 15.2)$  &  $(16.9 - 17.9)$ &  $(+0.4 - +1.0)$    &    $(2.5 - 4.2) \times 10^{-4}$ & $(1.0 - 1.4) \times 10^{4}$\\
\hline
\end{tabular}
\label{tab4}
\begin{tablenotes}
\item[] Columns 2 \& 3 are the redshift of the absorber and the $\HI$ column density, which are input parameters to\textsc{CLOUDY}. The subsequent columns list the total hydrogen column density, the abundance of carbon, the gas phase density range for a single phase solution, and the temperature of the gas predicted by the photoionization models. For the absorber at $z \sim 0.005$, there are two distinct absorbing components which are modelled separately.
\end{tablenotes}
\end{threeparttable}
\end{center}
\end{table*}

\subsection{Densities and temperatures for the $\mathbf{z_{abs}=0.00402}$ absorber towards SBS~$\mathbf{1122+594}$}
Unlike the previous absorber, the detection of different ionization stages of the same element in this absorber ($\CII$ \& $\CIV$, and $\SiIII$ \& $\SiIV$) allow us to constrain the density independent of the metallicity, or the $\HI$ column density and the uncertainties associated with it. The observed $\log~[N(\CII)/N(\CIV)] = -0.72~{\pm}~0.08$ is true for a density of $n_{\mathrm{H}} \sim (0.9 - 1.4) \times 10^{-4}~\cc$. At a similar density of $n_{\mathrm{H}} \sim (0.8 - 1.8) \times 10^{-4}~\cc$, the models also explain the observed $\log~[N(\SiIII)/N(\SiIV)] = -0.15~{\pm}~0.11$, with $\SiII$ as a non-detection. The metal lines are all thus consistent with a $n_{\mathrm{H}} \sim (0.9 - 1.5) \times 10^{-4}~\cc$ single phase origin. Unlike density, metallicity is poorly constrained from the models. At the lower limit of $\log~[N(\HI)~(\cmsq)] = 15.2$, the observed $\CII$, $\CIV$, $\SiIII$ and $\SiIV$ have a single phase origin at [C/H] $= 0.7$, and [Si/H] $= 0.3$ (see Figure~\ref{7}). At the other extreme, for the upper limit of $\log~[N(\HI)~(\cmsq)] = 17.3$, the abundances are as low as [C/H] = $-1.5$, and [Si/H] $= -1.8$. In this range, the models also predict $T = (1.3 - 2.4) \times 10^4$~K, a total hydrogen column density of $\log~[N(\mathrm{H})~(\cmsq)] = 18.1 - 20.6$, and line of sight thickness of $L = (3 - 1448)$~kpc. The lower limit on the absorber size is consistent with the diffuse CGM of a galaxy, whereas the upper limit is reminiscent of large scale sheets and filaments of the cosmic web linking massive halos, which are a few hundred kpc to several Mpc in dimension \citep{Bond2010,Gonzalez2010}. It is possible that the system resides in a region of several hundred kpc thickness constituting two or more merged halos. In such cases, one expects sub-solar metallicities in the absorbing gas.

\subsection{Densities and temperatures for the $\mathbf{z_{abs}=0.00574}$ absorber towards RXJ~$\mathbf{1230.8+0115}$}
The two clouds detected in the $z = 0.00574$ absorber towards RXJ~$1230.8+0115$ are modelled separately. The Cloud 1 and Cloud 2 are centered at $v(\CIV) = -34~\kms$ and $v(\CIV) = -2~\kms$ respectively. The ratio of column densities between $\CII$ and $\CIV$, and also between $\SiII$, $\SiIII$ and $\SiIV$, can be used to determine the gas density from the photoionization models. In both clouds, $N(\CII) > N(\CIV)$ indicating that the density has to be $n_{\mathrm{H}} > 2 \times 10^{-4}~\cc$. Similarly, $N(\SiII) < N (\SiIV)$, and $N(\SiIII)$, which are true for $n_{\mathrm{H}} < 5 \times 10^{-4}~\cc$. For Cloud 1, the observed $N(\CII)/N(\CIV) = 0.64~{\pm}~0.12$~dex occurs at a density of $n_{\mathrm{H}} = (4.2 - 5.5) \times 10^{-4}~\cc$. At a comparable density of $n_{\mathrm{H}} = (2.0 - 5.8) \times 10^{-4}~\cc$, the observed $N(\SiII)/N(\SiIII) = -0.70~{\pm}~0.13$~dex and $N(\SiII)/N(\SiIV) = -0.36~{\pm}~0.15$~dex can also be explained if the relative abundance of Si to C is $-0.2$~dex compared to solar, which is within the uncertainty introduced by the errors in the column densities of the C and Si ions. These ions can be attributed to a single phase medium with the density of $n_{\mathrm{H}} = (3.1 - 4.6) \times 10^{-4}~\cc$. Though the abundance pattern is consistent with being approximately solar, the uncertain $\HI$ column density results in a wide range of possible metallicities ($-1.8 <$~[X/H]~$< 0.3$) for this cloud (see Figure~\ref{8}). The models in this range suggest $T = (1.4 - 1.9) \times 10^4$~K, a total hydrogen column density of $\log~[N(\mathrm{H})~(\cmsq)] = 17.3 - 19.6$, and line of sight thickness of $L = (0.14 - 45)$~kpc.

For Cloud 2, the observed  $N(\CII)/N(\CIV) = 0.48~{\pm}~0.07$~dex is reproduced in the models at $n_{\mathrm{H}} = (3.8 - 4.7) \times 10^{-4}~\cc$. A comparable density is also obtained from the observed column density ratios between $\SiII$, $\SiIII$ and $\SiIV$, which are valid over the approximate range of $n_{\mathrm{H}} = (2.0 - 7.1) \times 10^{-4}~\cc$. A single phase solution requires $n_{\mathrm{H}} = (2.5 - 4.2) \times 10^{-4}~\cc$ and the relative abundance to be [Si/C]~$= -0.1~{\pm}~0.1$. The ionization models predict $T = (1.0 - 1.4) \times 10^4$~K, a total hydrogen column density of $\log~[N(\mathrm{H})~(\cmsq)] = 16.9 - 17.9$, and line of sight thickness of $L = (68 - 1073)$~pc. For the plausible range of $\HI$ column densities, the metallicity has to be a factor of 2 to 10 times higher than solar to explain the observed column densities of the metal lines. Such metallicities are atleast $\sim 0.5$~dex higher than the typical ICM metallicity obtained for the outskirts of clusters (and groups) from X-ray studies \citep{Mushotzky1978,DeGrandi2004,Werner2013,Tholken2016}.  

\section{SPATIAL DISTRIBUTION OF GALAXIES NEAR THE ABSORBERS} \label{Sec5}

\begin{figure*}
\begin{center}
        \includegraphics[width=550pt,trim=5cm 0.5cm 0cm 0.2cm,clip=true]{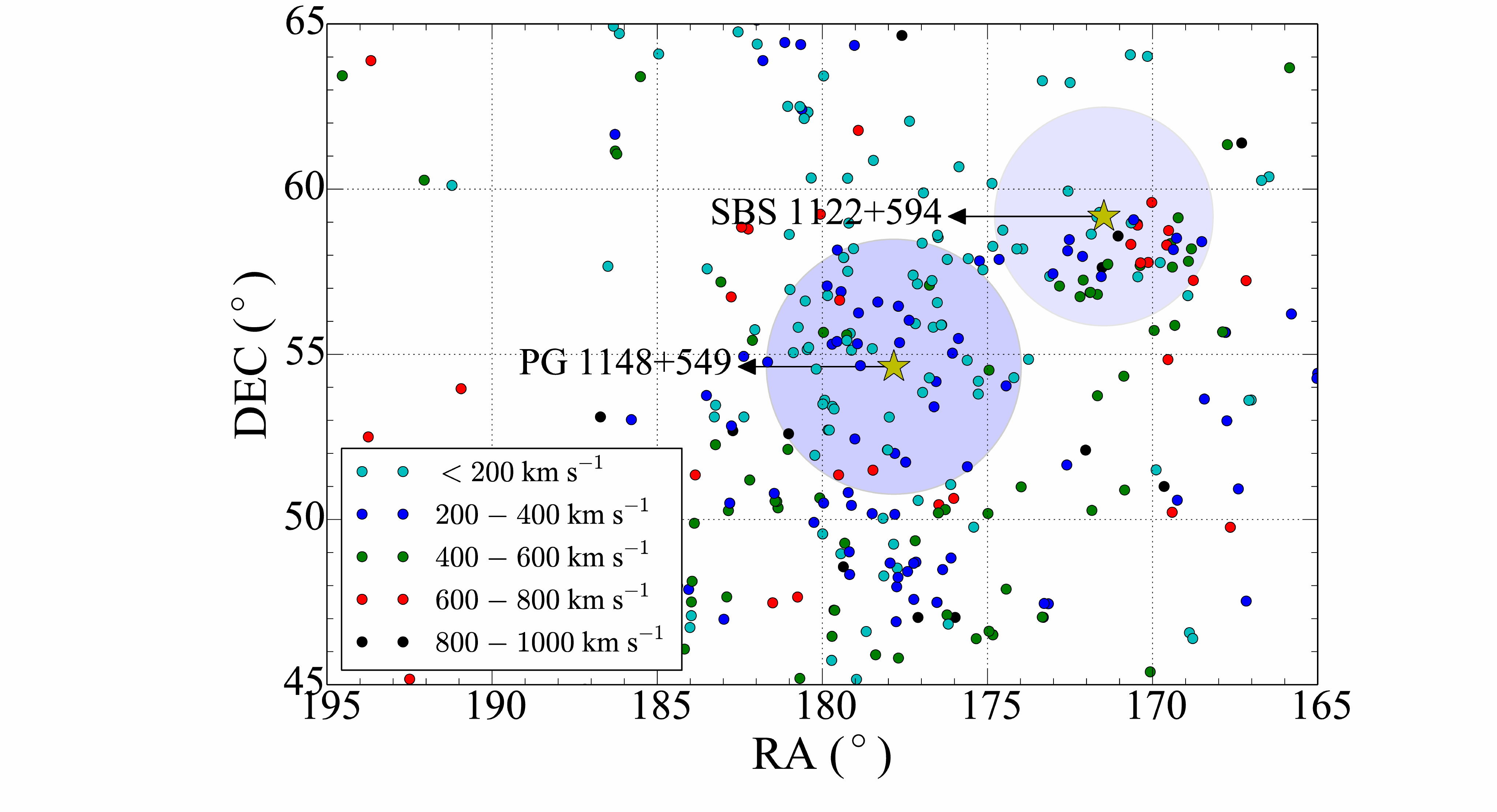}
        \includegraphics[width=510pt,trim=2cm 0.5cm 0cm 1.1cm,clip=true]{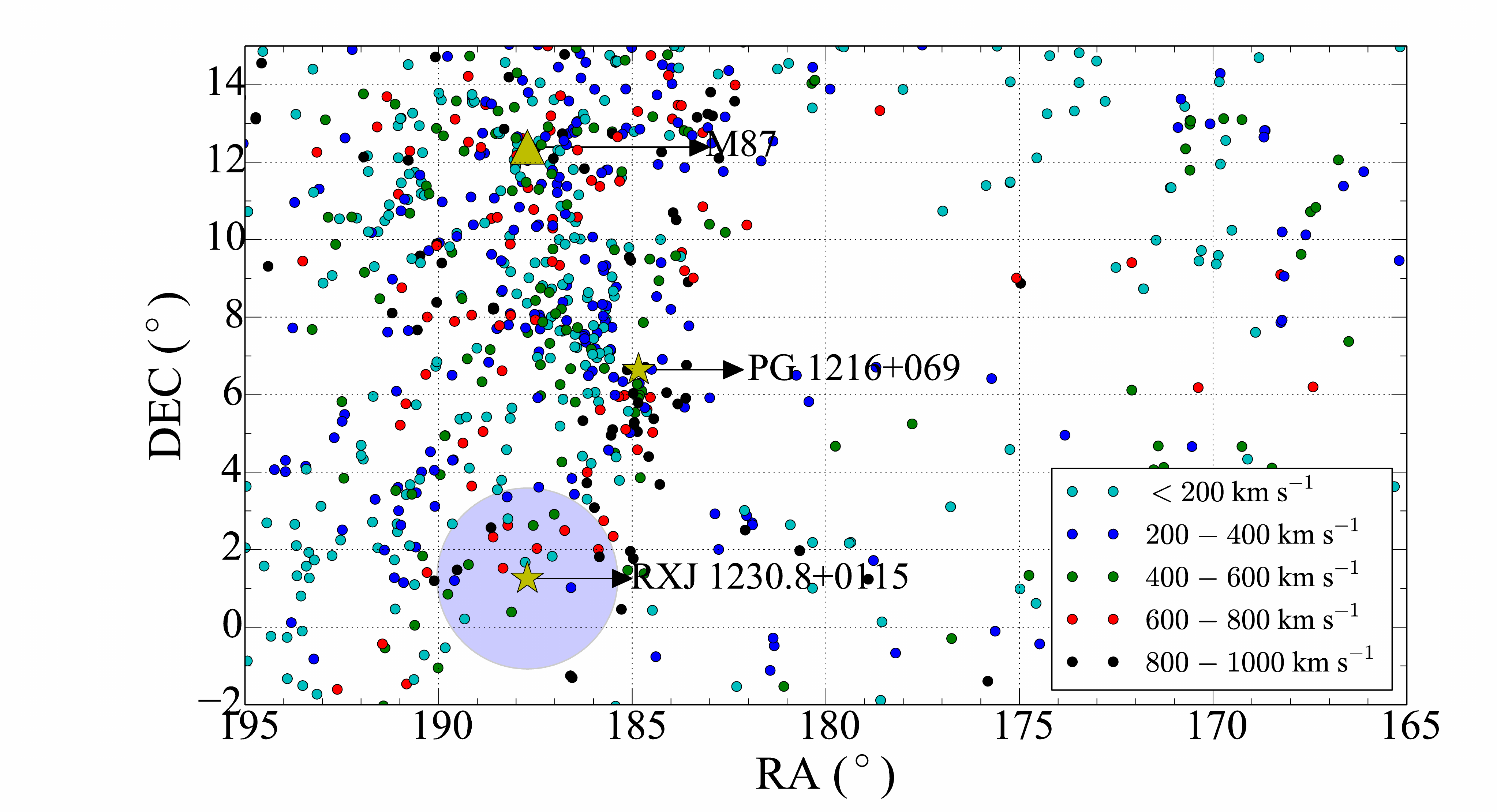}
    \caption{The plots show galaxy environments around the absorbers where the sightlines are represented with yellow stars and the galaxies with filled circles. The galaxies are color-coded to indicate their separation from the cluster center in intervals of 200~$\kms$ ranging from 0 to 1000~$\kms$. Regions within $\rho < 1$~Mpc of the absorber are represented as filled blue circles. The \textit{top} panel is the large-scale distribution of galaxies around $z_{abs}=0.00346$ and $z_{abs}=0.00402$ towards PG~$1148+549$ and SBS~$1122+594$ respectively. The \textit{bottom} panel shows distribution of galaxies around the $z_{abs}=0.00574$ absorber towards RXJ~$1230.8+0115$. The sightline traces the outskirts of the Virgo cluster with the cluster center, M$87$, indicated with a yellow triangle. The sightline towards PG~$1216+069$ has only $\HI$ (and no $\CIV$) at the redshift of the cluster.}
    \label{9}
\end{center}
\end{figure*}

\begin{table*} 
\caption{Galaxies within a projected separation of $\sim 1$~Mpc and $|\Delta v| = 600~\kms$ of the $z_{abs}=0.00346$ absorber towards PG~$1148+549$. The $z$ values are SDSS spectroscopic redshifts. $\Delta v$ is the systemic velocities of the galaxies with respect to the absorber. The error in velocity separation comes from the uncertainty in the spectroscopic redshift. The projected separation $\rho$ was calculated from the angular separation assuming a $\Lambda$CDM universe with parameters given in \citep{Bennett2014} using \citet{Wright2006}. Virial radii of the galaxies are calculated using the scaling relationship between $L/L^*$ and $R_{vir}$ given by \citet{Prochaska2011}. While determining the absolute magnitudes, appropriate K-corrections (which were minor) were applied using the analytical expression given by \citet{Chilingarian2010}. The Schecter absolute magnitude of $M^*_g = -20.18$ for the closest available redshift of $z = 0.07$ was taken from \citet{Ilbert2005}.}
\begin{center}
\small
\begin{tabular}{lccrcccrcc}
\hline
R.A.  &  Dec.   &   $z_{gal}$   &  $\Delta v$ ($\kms$)  &   $\rho$  (kpc)  &  $g$ (mag) & $g-r$ (mag) &  M$_g$ & $\rho/R_{vir}$\\
\hline
$178.84790$  &	$54.65734$  &   $0.00284$  &  	$-185~{\pm}~2$	&	 $152.1$	& 	$14.85$	& $0.4184$ & $-15.59$ & $1.41$\\
$178.48730$  &  $55.17149$  &   $0.00399$  &	$158~{\pm}~9$   &	 $171.6$   	&	$17.46$	& $0.515$ &  $-13.71$ & $2.25$\\
$176.76186$  &  $54.28802$  &   $0.00455$  &	$326~{\pm}~3$   &	 $183.9$   	&	$21.73$	& $-0.1194$ & $-9.73$ & $5.03$\\
$177.66223$  &  $55.35387$  &   $0.00317$  &    $-88~{\pm}~5$ 	&	 $190.4$        &	$13.43$	& $0.5425$ & $-17.24$ & $1.30$\\
$176.55608$  &  $54.17620$  &   $0.00332$  &    $-41~{\pm}~26$  &	 $225.4$        &   	$17.50$	& $0.483$ & $-13.27$  & $3.21$\\
$179.11719$  &  $55.12524$  &   $0.00371$  &    $75~{\pm}~5$    &	 $230.8$   	&	$12.08$	& $0.656$ & $-18.93$  &	$1.15$\\
$176.96400$  &	$53.84666$  &	$0.00340$  &	$-17~{\pm}~41$	&	 $241.2$	&   	$17.21$	& $0.4304$ & $-13.61$ & $3.22$\\
$178.93787$  &	$55.32074$  &   $0.00285$  &    $-181~{\pm}~6$  &	 $243.6$   	&	$13.11$	& $0.711$ & $-17.33$  &	$1.64$\\
$176.06174$  &  $55.03495$  &   $0.00476$  &    $388~{\pm}~4$   &        $285.2$  	&    	$14.43$	& $0.9703$ & $-17.13$ & $1.99$\\
$179.25780$  &	$55.41979$  &	$0.00403$  &    $170~{\pm}~3$	&	 $294.9$   	&	$16.47$ & $0.5519$ & $-14.73$ & $3.21$\\
$179.55705$  &  $55.38794$  &   $0.00323$  &	$-68~{\pm}~7$   &	 $323.2$   	&	$16.63$	& $0.6805$ & $-14.07$ & $3.97$\\
$179.15634$  &  $55.63325$  &   $0.00430$  &    $251~{\pm}~7$   &        $326.3$  	&    	$14.85$	& $0.7315$ & $-16.49$  & $2.57$\\
$179.26289$  &  $55.58678$  &   $0.00258$  &    $-264~{\pm}~10$ &        $336.8$  	&	$16.62$	& $0.5954$ & $-13.60$  & $4.52$\\
$179.70493$  &  $55.30689$  &   $0.00316$  &    $-89~{\pm}~28$  &        $329.3$  	&	$17.15$	& $0.5577$ & $-13.51$  & $4.49$\\
$175.61337$  &	$54.81902$  & 	$0.00415$  &	$206~{\pm}~0$	&	 $336.3$   	&	$15.06$	& $0.6248$ & $-16.19$  & $2.79$\\
$177.18156$  &  $55.92928$  &   $0.00356$  &    $31~{\pm}~0$    &        $351.3$   	&	$17.80$	& $0.5580$ & $-13.13$  & $5.14$\\
$180.18491$  &  $54.55421$  &   $0.00426$  &    $237~{\pm}~3$   &        $353.3$   	&	$16.98$	& $0.3441$ & $-14.33$  & $4.14$\\
$176.64185$  &  $55.82130$  &   $0.00360$  &    $41~{\pm}~3$    &        $356.5$   	&	$16.87$	& $0.3481$ & $-14.08$  & $4.37$\\
$175.87976$  &  $55.47895$  &   $0.00327$  &    $-57~{\pm}~5$   &        $364.9$   	&	$15.52$	& $0.3929$ & $-15.22$  & $3.63$\\
$176.61783$  &  $53.41206$  &   $0.00305$  &    $-121~{\pm}~4$  &        $365.1$   	&	$17.68$	& $0.4442$ & $-12.91$  & $5.56$\\
\hline
\end{tabular}
\label{tab5}
\end{center}
\end{table*}

\begin{table*} 
\caption{Galaxies within a projected separation of $\sim 1$~Mpc and $|\Delta v| = 600~\kms$ of the absorber towards SBS~$1122+594$.}
\begin{center}
\small
\begin{tabular}{lccrcccrrr}
\hline
R.A.  &  Dec.   &   $z_{gal}$   &  $\Delta v$ ($\kms$)  &   $\rho$  (kpc)  &  $g$ (mag) & $g-r$ (mag) &   M$_g$ & $\rho/R_{vir}$\\
\hline
$171.68462$  &	$59.15545$  &   $0.00402$  &  	$-1~{\pm}~2$	&	 $33.0$	        &	$14.15$	& $0.5476$  &  	$-17.03$ & $0.23$\\
$171.60812$  &  $59.29371$  &   $0.00415$  &	$38~{\pm}~0$   &	 $42.0$   	&	$17.07$	& $0.1584$  &	$-17.05$ & $0.50$\\
$170.64864$  &  $58.97792$  &   $0.00420$  &	$53~{\pm}~2$   &	 $141.1$   	&	$16.68$	& $0.2164$  &	$-14.17$ & $1.57$\\
$170.57457$  &  $59.07452$  &   $0.00524$  &    $363~{\pm}~3$ 	&	 $142.7$        &	$12.15$	& $0.6422$  &	$-14.62$ & $0.63$\\
$171.85620$  &  $58.63879$  &   $0.00418$  &    $47~{\pm}~3$   &	 $172.1$        &   	$16.95$	& $0.3424$  & 	$-19.63$ & $2.01$\\
$170.45935$  &  $58.94410$  &   $0.00564$  &    $483~{\pm}~2$   &	 $172.2$   	&	$17.41$	& $0.4058$  &	$-14.33$ & $1.95$\\
$172.52148$  &	$58.47168$  &	$0.00511$  &	$325~{\pm}~9$	&	 $267.9$        &   	$17.60$	& $0.4720$  &	$-14.52$ & $3.37$\\
$172.56021$  &	$59.94085$  &   $0.00341$  &    $-182~{\pm}~7$ &	 $285.7$   	&	$16.72$	& $0.3683$  &	$-14.11$ & $3.49$\\
$169.22469$  &  $59.13277$  &   $0.00561$  &    $473~{\pm}~7$   &        $348.9$  	&    	$18.06$	& $0.4847$  &	$-14.10$ & $4.46$\\
$172.57199$  &	$58.13391$  &	$0.00483$  &    $240~{\pm}~7$	&	 $358.4$   	&	$14.69$ & $0.5119$  &	$-13.86$ & $2.62$\\
$173.82558$  &  $58.88855$  &   $0.00346$  &	$-167~{\pm}~3$ &	 $375.8$   	&	$15.92$	& $0.4340$  &	$-16.90$ & $3.94$\\
$172.12175$  &  $57.96414$  &   $0.00331$  &    $-212~{\pm}~106$   &     $379.4$  	&    	$22.89$	& $1.6495$  &	$-14.93$ & $14.86$\\
$169.27298$  &	$58.51676$  & 	$0.00527$  &	$371~{\pm}~3$	&	 $397.3$   	&	$15.93$	& $0.3280$  &	$-7.79$ & $3.52$\\
$169.37073$  &	$58.16970$  & 	$0.00516$  &	$340~{\pm}~3$	&	 $448.7$   	&	$17.18$	& $0.3837$  &	$-15.85$ & $5.05$\\
$173.93172$  &	$58.19253$  & 	$0.00455$  &	$156~{\pm}~20$	&	 $486.9$   	&	$16.09$	& $-0.0283$  &	$-14.55$ & $4.72$\\
$174.53561$  &	$58.75829$  & 	$0.00417$  &	$44~{\pm}~2$	&	 $493.4$   	&	$13.67$	& $0.7474$  &	$-15.36$ & $3.17$\\
$169.77042$  &	$57.77752$  & 	$0.00415$  &	$38~{\pm}~0$	&	 $500.5$   	&	$17.67$	& $0.5158$  &	$-17.60$ & $6.74$\\
$174.11031$  &	$58.19138$  & 	$0.00404$  &	$4~{\pm}~5$	    &	 $509.5$   	&	$14.00$	& $0.4432$  &	$-13.58$ & $3.51$\\
$168.82320$  &	$58.19347$  & 	$0.00538$  &	$403~{\pm}~16$	&	 $511.1$   	&	$17.97$	& $0.3076$  &	$-17.21$ & $6.54$\\
$168.51042$  &	$58.41109$  & 	$0.00315$  &	$-261~{\pm}~53$	&	 $518.2$   	&	$23.36$	& $1.4792$  &	$-13.86$ & $22.43$\\
$171.55389$  &	$57.35352$  & 	$0.00486$  &	$248~{\pm}~2$	&	 $550.2$   	&	$16.38$	& $0.2660$  &	$-7.24$ & $5.48$\\
\hline
\end{tabular}
\label{tab6}
\end{center}
\end{table*}

\begin{table*} 
\caption{Galaxies within a projected separation of $\sim 1$~Mpc and $|\Delta v| = 600~\kms$ of the absorber towards RXJ~$1230.8+0115$.}
\begin{center}
\small
\begin{tabular}{lccrcccrrr}
\hline
R.A.  &  Dec.   &   $z_{gal}$   &  $\Delta v$ ($\kms$)  &   $\rho$  (kpc)  &  $g$ (mag) & $g-r$ (mag)  &   M$_g$ & $\rho/R_{vir}$\\
\hline
$187.76477$  &  $1.67564$  &   $0.00374$  &    $-598~{\pm}~9$    &	 $181.2$   	&	$21.77$	& $0.9497$ & $-9.26$  & $5.40$\\
$187.38655$  &  $0.83964$  &   $0.00752$  &    $527~{\pm}~4$    &	 $225.6$   	&	$19.00$	& $0.3489$ & $-13.55$  & $3.05$\\
$188.33652$  &  $1.52171$  &   $0.00556$  &    $-57~{\pm}~2$    &	 $292.0$   	&	$15.84$	& $0.3898$ & $-16.06$  & $2.49$\\
$187.46072$  &  $2.03156$  &   $0.00592$  &    $52~{\pm}~12$ 	&	 $348.6$	&	$17.29$	& $0.5523$ & $-14.75$  & $3.78$\\
$188.11647$  &	$0.39067$  &   $0.00505$  &    $-208~{\pm}~2$  &	 $409.9$   	&	$14.02$	& $0.3563$ & $-17.66$  & $2.60$\\
$186.58038$  &  $1.01962$  &   $0.00415$  &    $-476~{\pm}~0$   &        $493.7$  	&    	$16.64$	& $0.3234$ & $-14.61$  & $5.49$\\
$187.55627$  &  $2.62522$  &   $0.00544$  &    $-92~{\pm}~4$   &         $590.0$  	&    	$19.99$	& $-0.1524$ & $-11.84$  & $10.94$\\
$188.59177$  &  $2.32531$  &   $0.00587$  &    $36~{\pm}~4$   &          $593.9$  	&	$14.22$	& $0.531$ & $-17.80$  &	$3.68$\\
$188.21349$  &  $2.62828$  &   $0.00591$  &    $48~{\pm}~4$   &          $626.2$  	&	$20.47$	& $0.4347$ & $-11.56$ & $12.24$\\
$186.74348$  &	$2.49440$  &   $0.00564$  &    $-31~{\pm}~6$  &	         $672.2$   	&	$12.61$	& $0.6015$ & $-19.33$ & $3.14$\\
$187.06621$  &  $2.70083$  &   $0.00740$  &    $490~{\pm}~7$    &        $677.1$  	&    	$15.45$	& $0.4139$ & $-17.07$  & $4.79$\\
$188.64577$  &  $2.56881$  &   $0.00615$  &    $120~{\pm}~7$  &          $690.8$  	&    	$16.17$	& $0.4852$ & $-15.94$  & $6.02$\\
$187.01332$  &	$2.91377$  &   $0.00487$  &    $-261~{\pm}~3$	&	 $769.9$        &	$18.87$	& $-0.0573$  & $-12.73$  & $12.12$\\
$189.52155$  &  $1.47775$  &   $0.00616$  &    $121~{\pm}~5$   &         $782.2$  	&	$12.97$	& $0.6754$ & $-19.16$ & $3.76$\\
$189.59047$  &	$1.20210$  &   $0.00415$  &    $-476~{\pm}~3$  &	 $806.3$   	&	$17.71$	& $0.3707$ & $-13.54$ & $10.93$\\
$185.84760$  &  $1.81519$  &   $0.00630$  &    $162~{\pm}~4$    &        $832.1$  	&    	$17.06$	& $0.1353$ & $-15.10$  & $8.47$\\
$185.87500$  &  $2.00803$  &   $0.00605$  &    $90~{\pm}~9$  &           $848.6$  	&    	$17.17$	& $0.4142$ & $-14.91$  & $8.94$\\
$189.76031$  &	$0.84973$  &   $0.00532$  &  	$-129~{\pm}~4$	&	 $895.9$        &	$16.95$	& $0.3606$  & $-14.85$  & $9.55$\\
\hline
\end{tabular}
\label{tab7}
\end{center}
\end{table*}

\begin{figure*}
	\includegraphics[width=245pt,height=203.3pt]{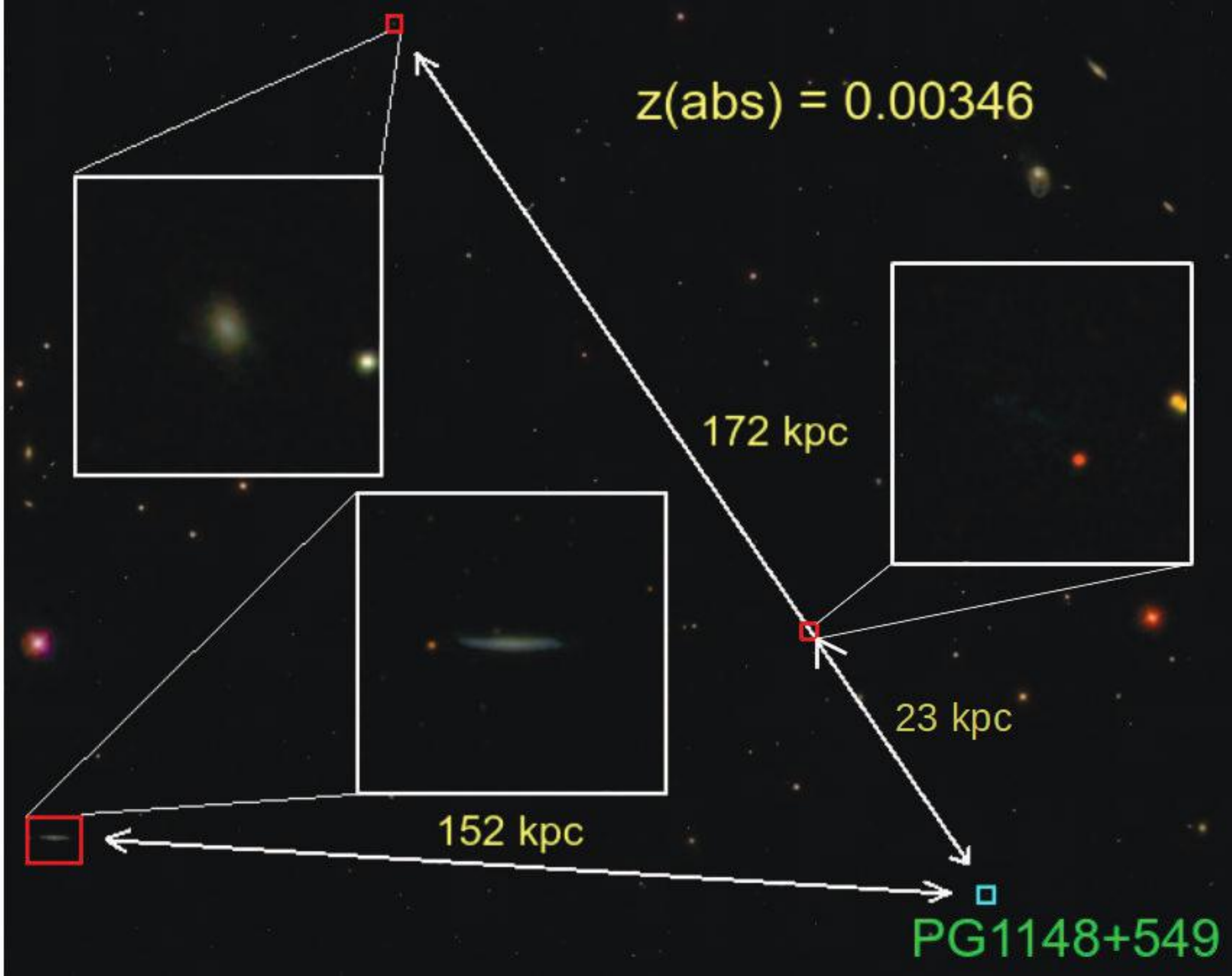}\quad \includegraphics[width=230pt,height=203.5pt]{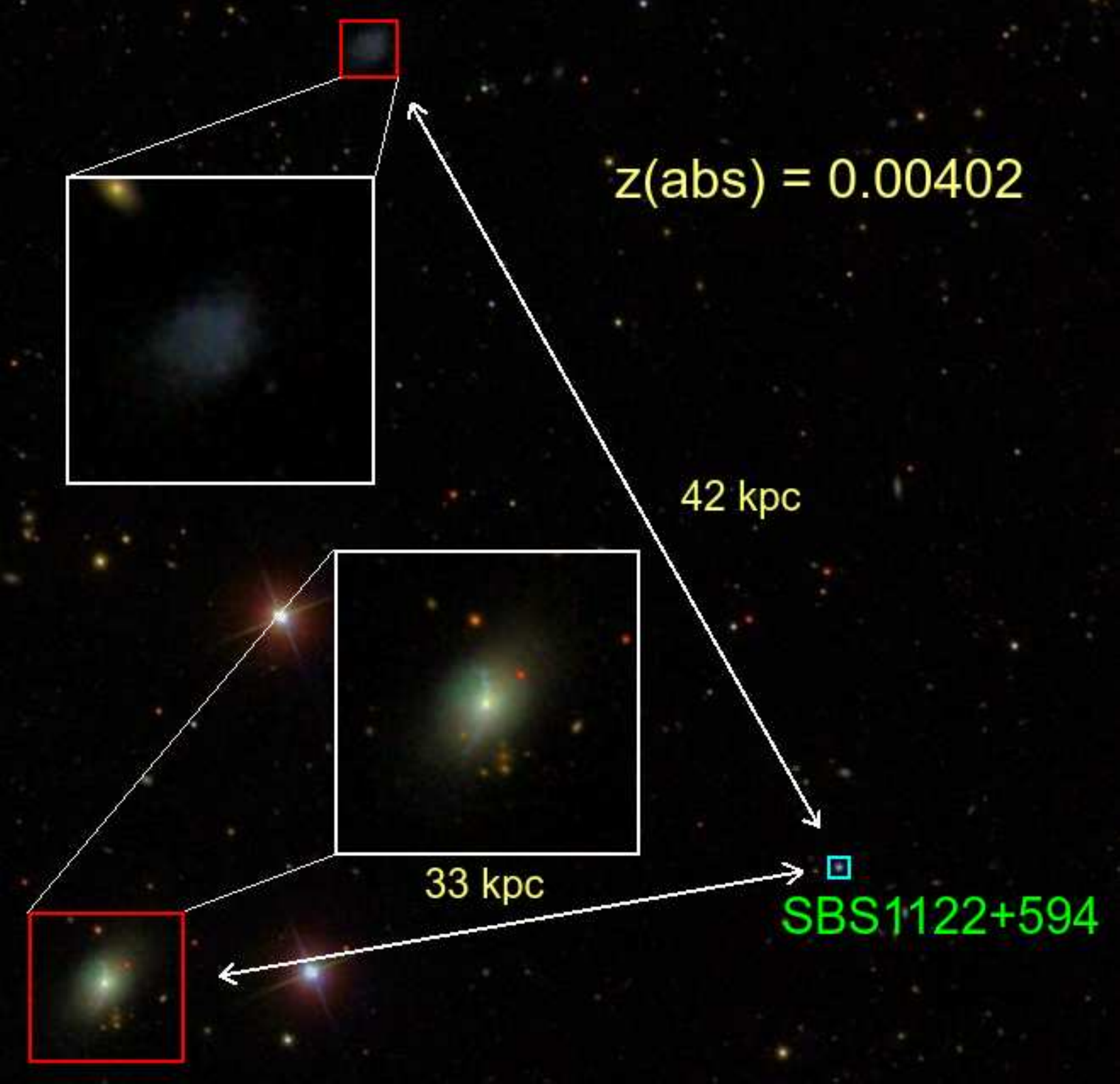}\quad \includegraphics[width=231pt,height=200pt]{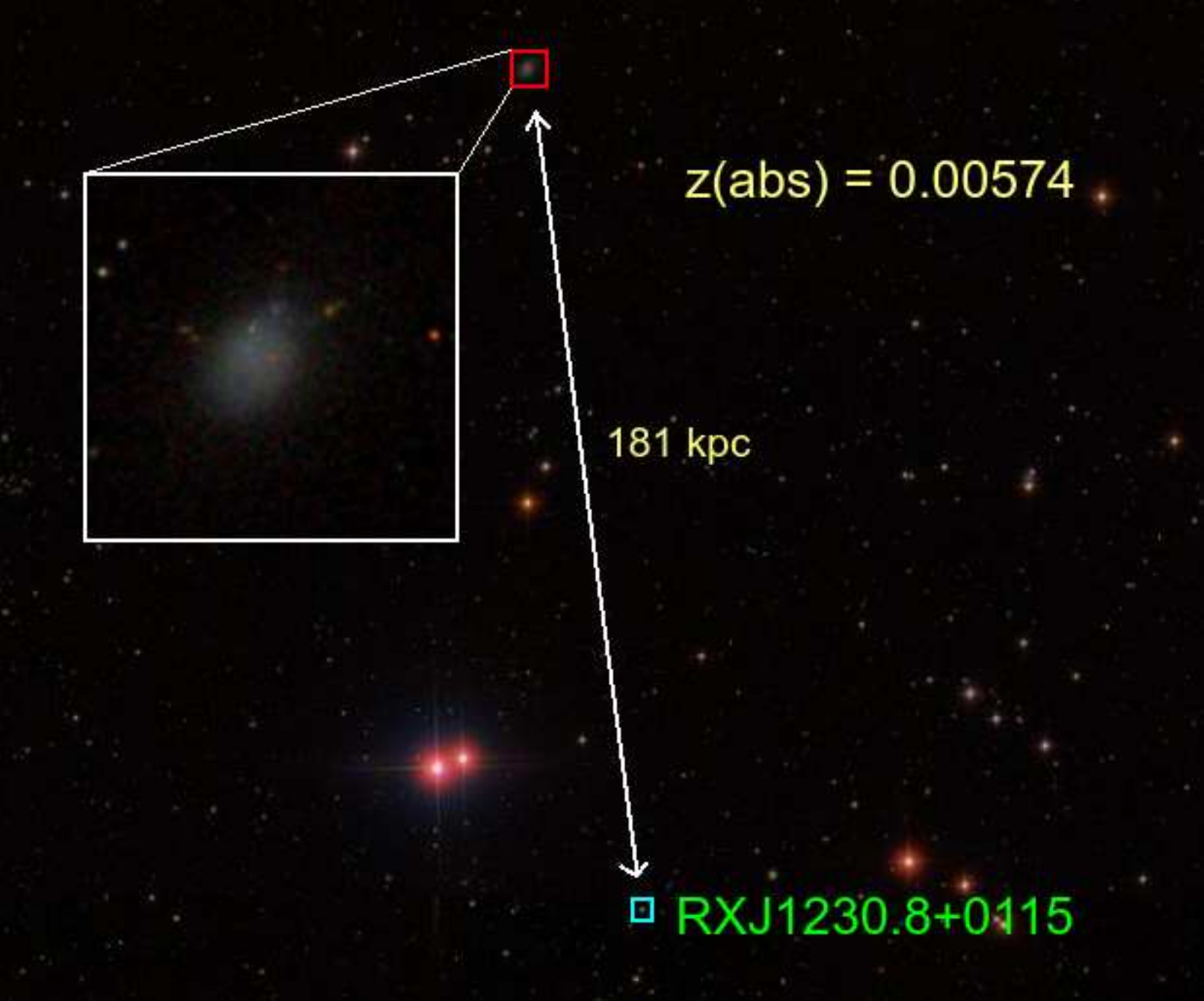}
    \caption{The SDSS $r$-band images of the field centered on the three quasars with the nearest galaxie(s) to the respective absorbers identified. The line of sight projected separation of each galaxy from the absorber is also indicated. The insets in each panel are zoomed-in versions of the galaxy images. In the top left panel, the dwarf galaxy at 23 kpc from the absorber towards PG~$1148+549$ (SDSS~J$115205.58+544732.2$) does not have a spectroscopic redshift given by SDSS. The redshift is provided in a much deeper survey conducted by \citet{Burchett2013}. The galaxy-absorber associations are discussed in Sec.~\ref{Sec5}.}
    \label{10}
\end{figure*}

The PG~$1148+549$ and SBS~$1122+594$ sightlines are separated from M87 by $12.4$~Mpc and $15.1$~Mpc, which are far out compared to the Virgo cluster radius of $\sim 2$~Mpc. SDSS shows the $z_{abs}=0.00346$ and $z=0.00402$ absorbers along these sightlines to be residing in local galaxy overdensity regions. 

The $z_{abs}=0.00346$ absorber has $89$ galaxies within a uniform projected separation of $1$~Mpc and $|\Delta v| \leq 600~\kms$. This is shown in Figure~\ref{9}. Such a large number of galaxies in a comparatively small volume of space suggests that the line of sight is probing a dense galaxy group or a poor cluster, based on the general attributes of clusters and groups given in \citet{Bahcall1999}. The median one-dimensional velocity dispersion of $\sigma = 183~\kms$ for the galaxies is more consistent with this being a rich group at a systemic velocity of $cz_{group} = 1065~\kms$. The $g-r$ color distribution of member galaxies and disk morphology apparent for some in the SDSS images imply this to be a spiral rich group \citep{Blanton2003} with a blue-to-red galaxy fraction of $12:1$.

The information on the 20 closest galaxies (by impact parameter) is given in Table~\ref{tab5}. Figure~\ref{10} shows the two bright galaxies UGC~$6894$ and LEDA~$2492981$ close by in impact parameter to the absorber, at $152$~kpc and $172$~kpc respectively. \citet{Burchett2013} have done a detailed analysis of galaxies in this field. They found, through deep imaging and spectroscopy, a dwarf irregular galaxy (SDSS~J$115205.58+544732.2$) unregistered in the SDSS spectroscopic database, at a much closer impact parameter of $23$~kpc (See Figure~\ref{10}). Using the galaxy's $g - r = 0.2$ color, stellar mass of $\sim 5 \times 10^5$~M$_{\odot}$ \citep{Burchett2013}, and the (M/L) scaling relationship for dwarf irregular galaxies \citep{Herrmann2016}, we estimate the galaxy's luminosity to be $L_g \sim 10^6 L_{\odot}$, which is consistent with its non-detection in the SDSS spectroscopic database. \citet{Burchett2013} conclude that the absorber is unlikely to be associated with this dwarf galaxy because of its large velocity separation ($\Delta v = +724~\kms$) with the absorber. On the other hand, UGC~$6894$ is at $|\Delta v| = 185$ km/s and $\rho = 152$~kpc ($1.4R_{vir}$) from the absorber. \citet{Burchett2013} infer the absorber to be a cool gas cloud accreted by UGC~$6894$. There is another dwarf galaxy, NGC~$3913$, close by in velocity to the absorber at $|\Delta v| = 88~\kms$ and $\rho = 190$~kpc ($1.3R_{vir}$). The $g - r = 0.542$ color makes it a blue galaxy \citep{Blanton2003} with an extended morphology seen in SDSS. We derive a star formation rate of SFR $= 0.001$~M$_{\odot}$~yr$^{-1}$ using the H$\alpha$ luminosity which is very low for it to be a dwarf starburst galaxy \citep{Martin2003}. A similar low SFR of $0.01$~M$_{\odot}$~yr$^{-1}$ is also estimated for UGC~$6894$. Interestingly, SDSS \citep[and also Table 2 of][]{Burchett2013} shows a sub-$L^*$ galaxy nearer in velocity and virial impact parameter. This galaxy (WR~$214$) is at $\rho = 230$~kpc ($1.2R_{vir}$) and $\Delta v = +75~\kms$ from the absorber. The galaxy has an extended morphology with an emission line dominated spectrum and $g - r = 0.656$ color, consistent with it being a blue galaxy \citep{Blanton2003}. However, the integrated luminosity in $H\alpha$ only suggests a star-formation rate of SFR $= 0.03$~M$_{\odot}$~yr$^{-1}$ \citep{Kennicutt1998}, which is much less compared to starburst galaxies in the local universe such as M82 \citep[$\sim 10$~M$_{\odot}$~yr$^{-1}$,][]{Mangano1978}. Thus, none of these galaxies are likely to be influencing absorption at large impact parameters from them through galactic-scale winds, though one cannot rule out the influence from past star-burst events. The sub-solar metallicity upper limit and the low densities of $n_{\mathrm{H}} \sim 10^{-5}~\cc$ are symbolic of cool intra-group gas. Such gas could also be in the process of getting accreted into the one of the nearby galaxies, as suggested by \citet{Burchett2013}. 
  
The $z_{abs}=0.00402$ absorber towards SBS~$1122+594$ has $51$ galaxies within an impact parameter of $1$~Mpc and $|\Delta v| \leq 600~\kms$, indicating a dense group environment \citep{Bahcall1999}. The mean velocity of the galaxies in the group is $cz_{group} = 1348~\kms$ with a velocity dispersion of $\sigma = 205~\kms$. The information on the $20$ closest galaxies (by impact parameter) is given in Table~\ref{tab6}. The group environment is dominated by blue galaxies as implied by their $g-r$ colors. Figure~\ref{10} shows the two (dwarf) galaxies closest to the absorber at projected separations of $33$~kpc and $42$~kpc. The galaxy at $33$~kpc is IC~$691$ which is at $0.2R_{vir}$ from the absorber. \citet{Keeney2006} have carried out a detailed analysis of this galaxy's association with the absorber. From the extinction corrected H$\alpha$ luminosity, they infer a SFR $\lesssim 0.24$~M$_{\odot}$~yr$^{-1}$ for IC~$691$ which makes it a dwarf starburst system. Based on estimates for the wind velocity and the galaxy orientation, \citet{Keeney2006} attribute the incidence of the $\CIV$ absorber to the starburst driven outflow from this dwarf galaxy. \citet{Keeney2006} obtain a metallicity of $-0.7$~dex for the galaxy, which is within the range of possible metallicities for the absorber given by the photoionization models. The other dwarf galaxy, at $42$~kpc of projected separation, is at $\rho = 0.5R_{vir}$ and $\Delta v = +37~\kms$ from the absorber. Using the integrated luminosity in $H\alpha$ determined from the SDSS spectrum of the galaxy, we obtain a SFR of $0.004$~M$_{\odot}$~yr$^{-1}$ \citep{Kennicutt1998}. Though this rate is too low for the galaxy to have enriched its CGM, the absorber could still be tracing the merged halos of the two galaxies, given their proximity in projected separation and line of sight velocity. 

Apart from the aforementioned possible associations, there is a spiral galaxy, NGC~$3642$, at $\rho = 142$~kpc ($0.6R_{vir}$), and $\Delta v = +362~\kms$, with ($L/L^*$)$_g$ $\sim$ $0.6$ and a star formation rate of $0.5$~M$_{\odot}$~yr$^{-1}$, indicated by its integrated $H\alpha$ luminosity. This galaxy is nearly face-on, with an inclination of $\sim 20.4^{\circ}$ \citep{Verdes2002} with respect to the plane of the sky. The $\CIV$ systems detected away from galaxies could be past outflows propagating through the galaxy's CGM or it could be left-over tidal streams from mergers \citep{Daigne2004,Songaila2006}. Indeed it has been proposed that the star formation in NGC~$3642$ is likely to have been induced by a merger with a gas-rich dwarf galaxy accreted from its local environment \citep{Verdes2002}. 

Rather than tracing any specific circumgalactic material, the absorber probably represents intragroup gas in the merged halos of these three galaxies which are all within one virial radii of the absorber. The relative chemical abundances of the gas can be influenced by outflows induced by star formation activity in IC~$691$ and/or the spiral galaxy NGC~$3642$, as well as from merger events and gas stripping of the CGM in the overall galaxy rich environment \citep{Chung2007,Yoon2013}. Such galactic scale events can lead to an increase in the covering fraction of $\HI$ and metals in the intergalactic regions \citep[e.g.][]{Hani2017}. Besides, environmental influences such as ram pressure stripping also act to remove gas from the CGM and redistribute it between the galaxies \citep{Yoon2013}. Given these, the absorber is more likely to be of intra-group origin rather than in the individual CGM of one of the nearby galaxies.

The $z=0.00574$ absorber towards RXJ~$1230.8+0115$ is near a subcluster within the Virgo cluster whose core region is occupied by M87. The absorber is at a projected separation of $4.1$~Mpc from M87 which is $2.6$ times the virial radius of the Virgo cluster as given by \citet{Yoon2012}, who identify this absorber as tracing gas along a filament in the outskirts of Virgo. The SDSS galaxy spectroscopic database shows $17$ galaxies within a projected separation (impact parameter) of $<$ 1 Mpc and $|\Delta v| \leq 600~\kms$ from the absorber as given in Table~\ref{tab7}. The number density of galaxies is consistent with this region being a subcluster or satellite group to Virgo. Beyond impact parameters of $\rho \lesssim 1.4R_{vir}$, it is unlikely for absorbers to be tracing individual galaxy halos \citep{Keeney2017}. The galaxies identified in the neighborhood of the absorber are (see Table~\ref{tab7}) well outside that range with the closest being at $\sim 2.5 R_{vir}$. With the SDSS spectroscopic database being nearly complete down to $0.001L^*$, it is safe to infer that the absorber is most likely probing cool ($T \sim 10^4$~K) intra-group gas rather than the isolated halo of a member galaxy of the group.  

However, the [C/H] $\geq 0.4$ for Cloud 2 and a possible [C/H] $\gtrsim 0$ for Cloud 1, obtained from the ionization modelling, requires that the absorber is tracing gas enriched by stars. Interstellar gas of near-solar metallicity could have been removed from one of the neighbouring galaxies through dynamical stripping, becoming part of the group medium. It is possible for galaxies to lose metal-rich gas through recurrent tidal forces and ram pressure stripping in dense cluster environments \citep{Chung2007,Tonnesen2007}. Such cool gas clouds are prevelant in the outer regions of hot X-ray emitting clusters \citep{Yoon2012,Yoon2017,Muzahid2017,Burchett2018}, in intra-group gas \citep{Bielby2017} as well as along large-scale intergalactic filaments \citep{Aracil2006}.  

The bottom panel in Figure~\ref{10} shows the nearest galaxy (CGCG~$014-054$) to this $z = 0.00574$ absorber, separated from it by $181$~kpc. The galaxy has a $g-r$ color of $\sim 0.95$, which makes it an elliptical galaxy \citep{Blanton2003} consistent with its SDSS broad band image. The galaxy is most likely a low mass elliptical with ($L/L^*$)$_g$ $\sim$ $4\times10^{-5}$. Tidal interactions and mergers between galaxies are expected to quench star formation \citep{Merritt1984,Abadi1999,Birnboim2003,Kerevs2005} leading them into the red sequence. However, the near solar or super-solar metallicity for a cloud in this system is higher than typical ISM metallicities in low mass galaxies, as both stellar and nebular metallicities is known to decrease with stellar mass. Thus, this nearest galaxy CGCG~$014-054$ may not directly account for the origin of the absorber. The next nearest galaxy is at a separation of $225$~kpc. To summarize, based on the available information on galaxies we associate the absorption system to metal-rich intragroup gas, with no conclusive hint on the source of the chemical enrichment.
 
\section{DISCUSSION \& SUMMARY}

Our analysis is primarily focused on establishing the ionization conditions, physical properties, and association with galaxies, for the three $\CIV$ absorbers at $z_{abs}=0.00346, 0.00402$ and $0.00574$ associated with the large scale environment around Virgo cluster. The absorbers are detected in the $HST$/COS spectra of PG~$1148+549$, SBS~$1122+594$ and RXJ~$1230.8+0115$ respectively. In all three instances, the metal line widths and ionization models are in accordance with the absorbers tracing cool ($T \sim 10^4 - 10^5$~K) and diffuse ($n_{\mathrm{H}} \sim 10^{-5} - 10^{-3}~\cc$) photoionized gas. The metallicities of these absorbers are less certain due to saturation in $\Lya$. 

There exists ambiguity in the literature on whether metal-line absorbers are associated with the halos of individual galaxies or with intra-group, intra-cluster medium. We have therefore tried to address the origin of these absorbers by looking at not just the nearest galaxies, but their large-scale distribution surrounding the absorbers. We found that all three absorbers reside in significant galaxy overdensity regions. The $z_{abs}=0.00574$ system, as known from earlier studies \citep{Yoon2012}, traces a sub-cluster in the outskirts of the Virgo cluster. However, unlike the Yoon et al. study which focused exclusively on $\HI$, the presence of metals along with $\Lya$ in these absorbers has allowed us to estimate the determine their temperature-density phase structure. The $z_{abs}=0.00346$ and $z_{abs}=0.00402$ systems probe dense galaxy groups in a region away from the Virgo cluster core. For these two latter absorbers, previous studies \citep{Burchett2013,Keeney2006} had only reported the nearest galaxies. Both these absorbers are consistent with origins in the respective cool phases of their intra-group medium. The key results from our analysis are summarized as follows: 

\begin{itemize}

\item The $z_{abs}=0.00346$ absorber towards PG~$1148+549$ traces a cloud with [C/H]~$< -0.2$. The four nearest galaxies to the absorber identified by SDSS and \citet{Burchett2013} are all at $\rho \sim (1 - 1.4) \rho_{vir}$ and with low SFRs of $\leq 0.03$~M$_{\odot}$~yr$^{-1}$. All of these galaxies are part of a large-group of $89$ galaxies found within $< 1$~Mpc and $\Delta v \leq 600~\kms$. The velocity offset between the absorber and the group is far less than the velocity dispersion of the galaxies within the group. Considering this, the low gas densities and sub-solar metallicities obtained from modelling, we hypothesize the absorber's origin to be in the cool ($T \sim 10^4$~K) photoionized phase of the intra-group gas.
  
\item The $z_{abs}=0.00402$ absorber towards SBS~$1122+594$ resides within the virial radii of three galaxies. The closest galaxy IC~$691$ is thought to contribute to the enrichment of the absorbing cloud through a starburst driven outflow \citep{Keeney2006}. The next closest galaxy is also a dwarf system. Given their proximity ($< R_{vir}$, $\Delta v < 100~\kms$) to the absorber, the line of sight could very well be intercepting the merged halos of both these dwarf galaxies. The third galaxy, NGC~$3642$, though within a virial radii, is at a larger velocity separation of $\Delta v = 363~\kms$ from the absorber. With no robust means to differentiate the CGM of a galaxy from its surrounding intergalactic space in dense galaxy environments, a circumgalactic or intra-group origin is equally likely for the absorber, though we favor the latter scenario. In either case, the $\CIV$ is tracing cool photoionized gas.   

\item The $z=0.00574$ absorber towards RXJ~$1230.8+0115$ is a metal-rich ([C/H]~$\geq 0.4$ for one of the clouds) system in the outskirts of the Virgo cluster. The metal-rich cloud could be tidally stripped interstellar gas from a faint low mass galaxy nearest to the absorber. There are $17$ galaxies within $1$~Mpc and $|\Delta v| < 600~\kms$, in agreement with \citet{Yoon2012} who identify this galaxy concentration as a subcluster to the Virgo. Since the nearest galaxy is at $\rho > 1.4R_{vir}$, the cool ($T \sim 10^4$~K) photoionized gas probed by this absorber is most likely dynamically stripped interstellar gas, now part of the group environment. The metallicity that we derive for Cloud 2 in this absorber is atleast $\sim 0.5$~dex higher than the typical ICM metallicity obtained for regions away from the core of clusters from X-ray studies.

\item In all three instances, the galaxy over-density regions associated with the absorbers are dominated by spirals. The $\HI$ - $\CIV$ absorbers thus seem to provide a means to track the multiphase reservoirs of gas in spiral-rich groups, extending the previous absorption line studies of similar environments to cooler ($T < 10^5$~K) gas phases \citep{Mulchaey1996,Stocke2014}.

\end{itemize}

\section*{ACKNOWLEDGMENTS}
We would like to sincerely thank the referee for a critical review which proved to be crucial for this study. We acknowledge the work of people involved in the designing, construction and deployment of the COS onboard $HST$. We also wish to extend our thanks to all those who had carried out data acquisition through Far-UV observations towards the sightlines mentioned in this paper. The plots in Figs~\ref{2},~\ref{3} and~\ref{9} were generated using the graphics environment developed by \citet{Hunter2007}.


\bibliographystyle{mnras}
\bibliography{citations}

\bsp	
\label{lastpage}
\end{document}